\documentstyle[prl,aps,preprint,eqsecnum,epsfig]{revtex}
\tightenlines
\newcommand{\bfrac}[2]{\displaystyle{\frac{\displaystyle #1}
{\displaystyle #2}}}

\begin{document}
 
\title{Perturbative Approach to the Quasinormal Modes of Dirty Black Holes}
 
\author{P.\ T.\ Leung${}^{(1)}$, Y.\ T.\ Liu${}^{(1)}$,
W.\ M.\ Suen${}^{(1, 2)}$,  C.\ Y.\ Tam${}^{(1)}$  and K.\ Young${}^{(1)}$}
 
\address{${}^{(1)}$Department of Physics, 
The Chinese University of Hong Kong, Hong Kong, China}
\address{${}^{(2)}$McDonnell Center for the Space Sciences, Department of 
Physics, Washington University, \\
St Louis, MO 63130, U S A}
 
\date{\today}
 
\maketitle

- 
\begin{abstract}
 
Using a recently developed perturbation theory for uasinormal modes
(QNM's), we evaluate the shifts in the real and imaginary parts of the
QNM frequencies due to a quasi-static perturbation of the black hole
spacetime.  We show the perturbed QNM spectrum of a black hole can have
interesting features using a simple model based on the scalar wave
equation.

\end{abstract}
 
\draft

\pacs{PACS numbers: 04.30.Db}
 
\section{Introduction}  

The observational consequences of black holes interacting with their
astrophysical environments have been a subject of much interest for
the last 30 years.  Within the next few years, it is expected that the
new generation of gravitational wave observatories (LIGO, VIRGO)
\cite{detector} will be able to detect gravitational waves emitted by
black holes excited by matter, or even other black holes, falling into
them.  It has long been known that the gravitational waves emitted in
such a process will carry a signature associated with the well-defined
quasinormal mode (QNM) frequencies of the black hole \cite{qnm}, and
will, among other things, provide confirmation of the existence of
black holes.  Numerical simulations \cite{num} suggest that in some
cases the QNM ringing may even dominate the signal.

QNMs of black holes have been extensively studied with the black hole
perturbation theory \cite{chand}.  If a black hole settles down in an
otherwise empty and asymptotically flat spacetime at the end-point of
dynamical evolution, it will be a Kerr black hole (Schwarzschild
hole in the case of zero angular momentum) \cite{nohair}.  Weak (linearized)
gravitational waves propagating on the Kerr or Schwarzschild
background can be described by the Klein-Gordon equation \cite{chand}:

---kg
\begin{equation}
\left[ {\partial_{t}^{2}} - {\partial_{x}^{2}} +V(x) \right] \Phi(x,t) = 0~,
\label{eq:kg}
\end{equation}
where $x$ is a radial coordinate, $\Phi$ is the radial part 
of a combination of the linearized changes in the metric
functions representing the gravitational wave,
and the outgoing wave boundary condition is appropriate
for waves escaping to infinity.  The potential $V(x)$ describes the
scattering of the gravitational waves by the background
geometry.  For example, in the case of a Schwarzschild hole of mass $M$, 
$V$ is the Regge-Wheeler potential \cite{chand,rw}:
---schw
\begin{equation}
V(x) = \left( 1 - \frac{2M}{r} \right)
\left[ \frac{l(l+1)}{r^2} + (1 - s^2) \frac{2M}{r^3} \right]
\label{eq:schw}
\end{equation}
for each angular momentum sector $l$, 
where $x = r + 2M \ln \left(r/2M -1 \right)$, $s$ is the spin of the field 
($s=2$ for gravitational waves), and $r$ is the circumferential radius. 

A single-frequency solution [$\Phi \propto \exp(-i \omega t)$] with
the outgoing wave boundary condition is a QNM, with Im $\omega < 0$.  The QNM
spectra of Kerr and Schwarzschild black holes have been extensively
studied theoretically \cite{bholea} and numerically
\cite{others,bholen}, and the positions of the QNM frequencies in the 
$\omega$-plane are known in detail \cite{others,bholen}. 

\begin{figure}
\centerline{\epsfig{file=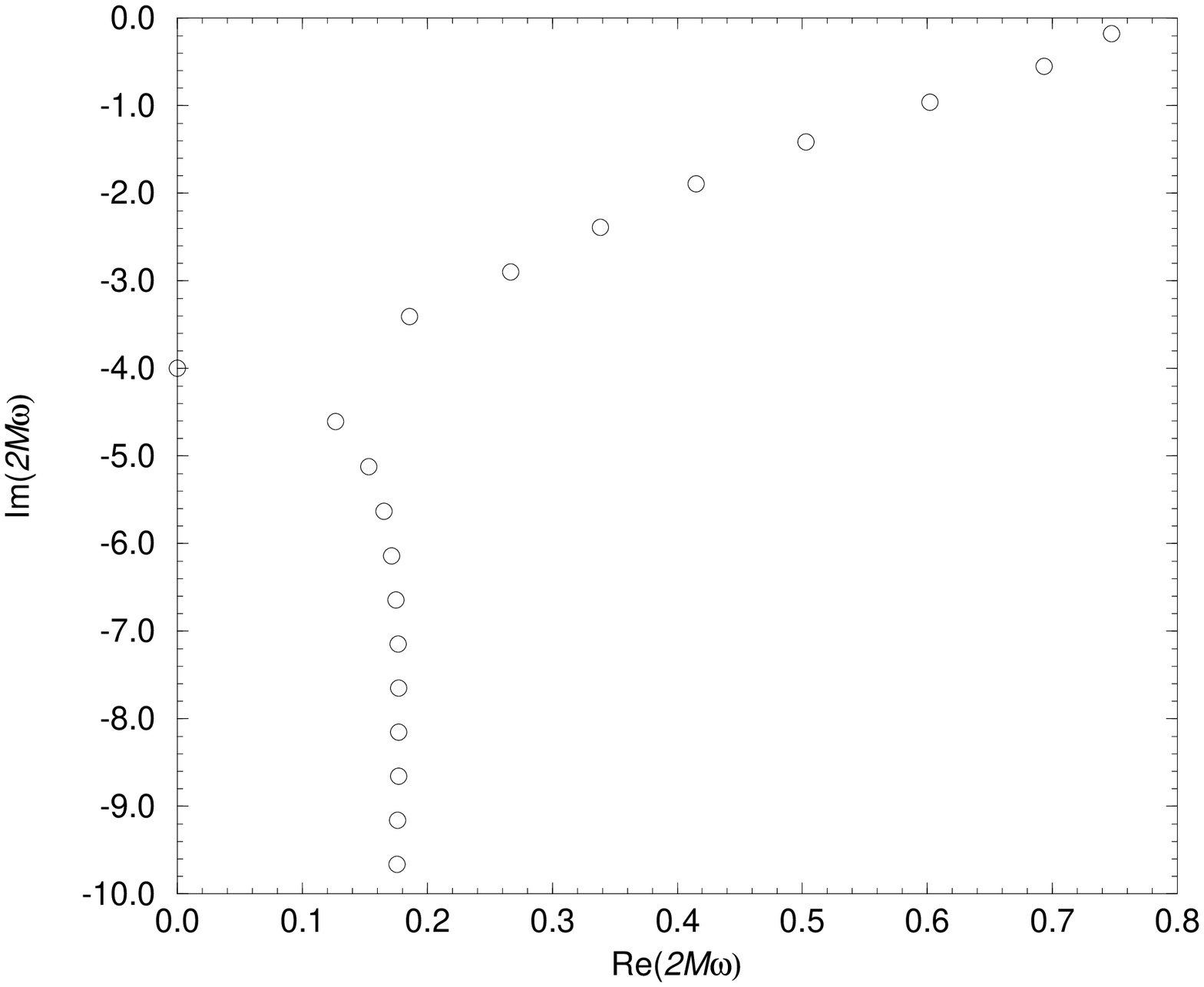,height=15cm,angle=270}}
\vskip 2mm
\caption{The distribution of QNMs of a Schwarzschild black hole for $l=s=2$.}
\label{fig:one}
\end{figure}

For each given $l$ in (\ref{eq:schw}), the QNMs extend downwards in a
string in the $\omega$-plane, with Re $\omega$ nearly constant and Im
$\omega$ nearly uniformly spaced \cite{others} (\mbox{Fig.\
\ref{fig:one}}).  The known pattern of frequencies provides a template
against which one can try to determine the nature of the source.  For
an isolated black hole, the no-hair theorem \cite{nohair} implies that
the spectrum is described by only two parameters, the mass $M$ and the
angular momentum $J$ of the hole.  However, the black holes that are
likely to be observed will not be isolated, but will likely be
situated at the centers of galaxies, or will be surrounded by massive
accretion disks.  Therefore the observed spectra should not be matched
against those of a pure Kerr or Schwarzschild black hole, but to one
perturbed by its surrounding --- a {\it dirty} black hole.  We should
immediately caution the reader that while gravitational waves from
black holes are expected to be detected within the next few years, a
determination of the QNM spectrum with the frequencies of a few modes
included might not be an easy task.  Indeed to what extent the
gravitational radiation from {\it realistic} black hole events would
be dominated the QNM spectrum is still a matter of much controversy.
However, in as much as the goal of the gravitational wave observatories is
to obtain astrophysical information of our universe (the ``O'' in
``LIGO''), there is no doubt that we will eventually have to face this
problem of the QNM spectra of {\it dirty} black holes.

We note that two kinds of perturbations are involved here.  In the
standard black hole perturbation theory \cite{chand}, (\ref{eq:kg}) is
obtained by linearizing the metric about the Kerr or Schwarzschild
background, and the time-independent eigenvalue problem (with the
outgoing wave boundary condition) determines the QNM spectra of
isolated holes.  The second type of perturbations are {\it the
perturbations that change the background} on which the wave
propagates, e.g., by the presence of an accretion disk near the black
hole.  The perturbation of the background can often be regarded as
quasi-static, and hence separable from that of the gravitational wave
perturbation by the time scales involved (in a suitable gauge choice).
In this paper we focus on time-independent perturbation of the
background described by (\ref{eq:kg}) with a potential $V(x) = V_0(x)
+ \mu V_1(x)$, $|\mu| \ll 1$ (a model problem is given in \mbox{Sect.\
\ref{sect:shell}} below).  We are led to study the QNM frequencies of
the following eigenvalue problem in powers of $\mu$:

---kg1
\begin{equation}
- \phi''(x) + \left[ V_0(x) + \mu V_1(x) \right] \phi = \omega^2 \phi~.
\label{eq:kg1}
\end{equation}

(\ref{eq:kg1}) is appropriate for considering the
perturbed Klein-Gordon wave equation describing the propagation of
scalar waves in a gravitationally perturbed (dirty) black hole
spacetime (as will be shown below).  In this paper we show how the
disturbed QNM spectrum can be determined for such a system.  This
represents the first step towards determining the disturbed
gravitational wave QNM spectrum of a dirty hole, the real case of
physical interest.  For a realistic black hole perturbed by an
external matter fluid source, the gravitational wave QNM spectrum also
involves the fluid modes.  Just as in the case of the perturbation of
fluid star, we would expect two types of perturbations, with one
strongly involving the fluid motion (e.g., the polar $f, g, p$
modes), while the other only weakly involving the fluid (e.g., the axial modes).
We expect the scalar perturbation studied in this paper to be more
easily generalizable to the latter kind of modes.  A more complicated
set of equations would have to be used to describe the former type of
perturbation involving the ``fluid" modes in the shell of matter
outside the hole.  While the calculation in this paper may not be
easily generalizable to these more complicated situations involving
gravitational waves coupled to matter, we note that a full
perturbation treatment of the gravitational wave case will not be
possible without a thorough understanding of the behavior of the
Klein-Gordon wave equation with a perturbed potential, namely the
system studied in this paper.

While the perturbed Klein-Gordon wave equation (\ref{eq:kg1}) is
superficially similar to standard textbook problems, e.g., the usual 
Rayleigh-Schr\"{o}dinger perturbation theory (RSPT),
we note that the perturbation problem encountered here
is fundamentally different: 
the outgoing wave condition renders the system 
physically nonconservative (energy escapes to infinity) and the associated
operator $[ - d^2/dx^2 +V(x) ]$ non-hermitian; hermiticity underpins  
the usual RSPT.  

The difficulty coming from the non-hermiticity can be seen in several
guises if one tries naively to transcribe the usual formulas.  Unlike
the hermitian case, now the unperturbed eigenstates do not in general
form a complete set for expansions \cite{kg-comp,ching-comp}, at least not in 
the case of black holes.  The usual RSPT formula in terms of a sum over
intermediate eigenstates is therefore inapplicable.  Even the
first-order shift, which does not involve a sum over intermediate
states, cannot be given by the usual formula $\langle \phi_0 | \mu V_1
| \phi_0 \rangle / \langle \phi_0 | \phi_0 \rangle$, in obvious
notation --- the usual inner product leads to $\langle \phi_0 | \phi_0
\rangle = \mbox{\Large $\int$}_{-\infty}^{\infty} dx \phi_0^* \phi_0 =
\infty$ since a QNM wavefunction extends over all space (and indeed grows
exponentially at infinity).

So far, the perturbation of black hole QNMs
has attracted little attention, partly because a perturbative
formalism for the QNMs of an open system, as opposed to the normal modes (NMs)
of a conservative system, has not hitherto been available.
In this paper we develop such a formalism, which then
opens the way to
extracting information about the astrophysical environment of the black holes
from the observed signal, beyond the mass and the angular momentum of the hole.

This paper is a follow-up of \cite{dirtyhole.prl}, which outlined some
of the results derived in this paper. In \mbox{Sect.\ \ref{sect:form}}
of this paper, we develop a formulation for the perturbation of QNM
systems.  As a first step in this direction, we limit ourselves to the
scalar wave case, in which the evolution is described by a single
Klein-Gordon equation with a perturbed potential.  The shifts in both
the real and imaginary parts of the QNM frequencies $\omega$ are
obtained in quadratures in terms of $\mu V_1(x)$, in principle to
arbitrary order in $\mu$.  Given the precision of the observational
data that is possible in the near future (indeed at this point it is
not clear how many QNM's one can extracted from the waveforms of black
hole events, given the S/N of even the advanced phase LIGO), the
emphasis is on the first-order shift.  The shifts when expressed in
terms of a generalized inner product take a form similar to that in
RSPT. The perturbative results for a Schwarzschild black hole are
derived in \mbox{Sect.\ \ref{sect:genprop}}.  We show that a function
$H(x)$ can be defined which depends {\it only} on the original {\it
unperturbed} system.  We investigated and presented in detail the
properties of this function $H$ for the black hole case.  This
function controls the phase and magnitude of the first-order shift of
the spectrum for {\it any} given perturbation (not just for the model
problem in this paper), hence providing insight to the properties of
the black hole spectra in general.  \mbox{Sect.\ \ref{sect:shell}}
illustrates these results with a model problem where the perturbation
is due to a spherical shell of matter located at a fixed radius, and
we study scalar wave propagating in this background.  The
frequency shifts are obtained using the perturbation formula, and
compared to numerical results.  We show even in this simple case the
spectra for shells located at different radii contain very interesting
features.  These features can be understood by the perturbation
formula, demonstrating the power of the perturbation formula in
providing understanding of the perturbed spectra.

-------
\section{Formulation}\label{sect:form}

Waves defined by differential equations such as (\ref{eq:kg1})
or the optical analogs described by
the wave equation \cite{waveeq} are open systems if the 
outgoing wave boundary condition is imposed;
as such they are physically nonconservative and mathematically
non-hermitian.  The usual tools of mathematical physics based
on the hermiticity of the defining operator would not, in general,
be expected to apply.  In attempting to develop generalizations of the
familiar formalisms for conservative systems, it is important to recognize that
these open systems fall into two broad classes.
In the first, the potential $V(x)$ has discontinuities
at $x= a_1, a_2$, and vanishes at infinity suitably fast
 (``no tail''); when these conditions are satisfied, the QNMs form a
complete set on the interval $[a_1,a_2]$, and the usual formalism
can be carried over with minimal changes \cite{kg-comp,two-comp}.
This is in itself somewhat surprising, in that the dynamics are controlled 
entirely by the resonances without the need to add any ``background".
However, when the discontinuity or ``no-tail'' conditions are not
satisfied --- and this is the situation of interest in this paper
--- the QNMs are in general not complete, and the dynamics is not completely
controlled by the QNMs.  The QNMs are manifested in the complex frequency plane
as poles of the Green's function; now, there is in addition a cut 
along the negative $\mbox{Im } \omega$ axis
\cite{kg-tail}.  The cut arises from scattering by the asymptotic part
of the potential (in the sense that if $V(x)$ is truncated at any finite
distance, the cut disappears).  This is of course the case for the radial problem of
a black hole, where $V(x)$ has no discontinuity and goes
asymptotically as a centrifugal barrier plus $\sim \log x /x^3$; this
tail generates the cut in the complex frequency plane, extending all the way
to $\mbox{Im } \omega \rightarrow 0$, and consequently
leads to the long time power-law behavior $\sim t^{-\alpha}$ of
the dynamics \cite{kg-tail,late-time}.  The point to be stressed here is
that since the QNMs are not complete, a
perturbation formalism based on a sum over intermediate states will
not be appropriate.

In this paper we need to rely on a different approach, which does not
require a complete set of QNMs, by generalizing
the logarithmic perturbation theory (LPT) 
\cite{went,price,polik,aa1,zero1,3d,zero2,lpt},
which focuses on the logarithmic derivative
$f(x)=\phi'(x) / \phi(x)$.  Three features
are relevant in the present context.
First, the numerical determination of QNMs is known 
to be difficult, essentially because the exponentially growing
solutions are numerically sensitive; this makes it important
(indeed more important than would be the case for NMs)
to develop analytic techniques, as well as semi-analytic techniques
such as perturbation theory.
Second,
LPT does not require a complete set of unperturbed
states for expansion, and is therefore well suited to situations
where QNMs are not complete, or circumstances where only a few QNMs are known
and others are not.  (This is in practice often the case, since
QNMs with large $- \mbox{Im } \omega$ are difficult to
obtain, even numerically. The case of gravitational waves
propagating away from a black hole is precisely of this type.)
Thirdly, QNMs, being complex,
are not in general plagued by nodes for real $x$, where $f(x)$ would be 
singular.  In the case of NMs, the zeros of excited
state wavefunctions require special treatment \cite{zero1,zero2};
fortunately for QNMs, this is not a problem.

In terms of $f(x)$, (\ref{eq:kg1}) becomes the Riccati equation

---ricc
\begin{equation}
f'(x) + f^2(x) - \left[ V_0(x) + \mu V_1(x) \right] + \omega^2 = 0~.
\label{eq:ricc}
\end{equation}
The function $f(x)$ satisfies {\it both} the left boundary condition
($f(x) \rightarrow -i\omega$ as $x \rightarrow -\infty$) and the right
boundary condition ($f(x) \rightarrow +i\omega$ as $x \rightarrow
+\infty$), with $\omega$ in (\ref{eq:ricc}) being the eigenvalue
$\omega$.  However, for any $\omega$ (whether or not an eigenvalue),
we can define two solutions $f_{\pm}(\omega,x)$ by the boundary
conditions $f_{\pm}(\omega,x) \rightarrow \pm i\omega$ as $x
\rightarrow \pm \infty$.  At an eigenvalue $\omega$, $f_+(\omega,x) =
f_-(\omega,x) = f(x)$. We expand the eigenvalue $\omega$ and the
eigenfunction $f$ in powers of $\mu$:

---omega
\begin{equation}
\omega = \omega_0 + \mu \omega_1 + \mu^2 \omega_2 + \cdots
\label{eq:omega}
\end{equation}

---f
\begin{equation}
f \equiv f_0 + g = f_0 + \mu g_1 + \mu^2 g_2 + \cdots~,
\label{eq:f}
\end{equation}
where $f_0$, assumed known, satisfies the Riccati equation
(\ref{eq:ricc}) with the potential $V_0$ and frequency $\omega_0$.

Putting (\ref{eq:omega}) and (\ref{eq:f})
into the Riccati equation (\ref{eq:ricc}) and upon comparing powers of $\mu$,
one finds, in a straightforward manner, that

---gn
\begin{equation}
g_n' + 2f_0g_n + 2\omega_0 \omega_n = V_n
\label{eq:gn}
\end{equation}
for $n = 1, 2, \cdots$, in which $V_1$ is the perturbing potential in 
(\ref{eq:kg1}), and $V_n$, $n>1$, is  
the effective $n$th order potential, defined in terms of a combination of lower-order 
quantities:

---vn
\begin{equation}
V_n(x) = -\sum_{i=1}^{n-1} \left[ g_i(x) g_{n-i}(x) + \omega_i \omega_{n-i} 
\right]~.
\label{eq:vn}
\end{equation}

\noindent
Using the integrating factor $\exp [2\int dx f_0(x)]$, (\ref{eq:gn}) can be 
solved; but since the resultant $g_n$, related to an
eigenfunction, must satisfy
{\it two} boundary conditions, this imposes a condition on the remaining
free parameter $\omega_n$.  The details of the derivation
will be given elsewhere \cite{lpt}.
Ignoring convergence problems for the moment, and noting that
the logarithmic derivatives at spatial infinity are unaffected by the
perturbation, by definition of the boundary conditions,
one obtains the following formal expressions,
which constitute the core of LPT \cite{fn1}:

--result
\begin{equation}
\omega_n = \frac{ \langle \phi_0 | V_n | \phi_0 \rangle} 
{ 2 \omega_0 \langle \phi_0 |  \phi_0 \rangle }~.
\label{eq:result}
\end{equation}
Here we have introduced the suggestive notation

---matelm
\begin{equation}
\langle \phi_0 | V_n | \phi_0 \rangle  =
\int_{-\infty}^{\infty} V_n(x) \phi_0^2(x) dx
\label{eq:matelm}
\end{equation}

---norm
\begin{equation}
\langle \phi_0 | \phi_0 \rangle =
\int_{-\infty}^{\infty} \phi_0^2(x) dx~.
\label{eq:norm}
\end{equation}

These results express the $n$th-order correction to the eigenvalue in 
quadrature in terms of lower-order quantities. The $n$th-order correction to 
the logarithmic derivative is given by 

---function
\begin{equation}
g_n(x) = \left\{ \int_{-\infty}^x dy \left[ V_n(y) - 2\omega_0 \omega_n \right]
 \phi_0^2(y) \right\}  \phi_0^{-2}(x)~.
\label{eq:function}
\end{equation}
When $g_n$ is substituted back in (\ref{eq:vn}) to obtain $V_{n+1}$,
the cycle of iteration is complete, and one can in principle obtain the 
corrections to any order.  The strategy outlined here is basically the same as in the
conventional LPT for hermitian systems.

The above expressions are formal and suggestive,
but hide an essential problem: because the unperturbed wavefunctions
go as $e^{\pm i\omega_0 x}$ as $x \rightarrow \pm \infty$, the integrands
contain factors $\phi_0(x)^2 \propto e^{2\gamma_0 |x|}$ as 
$|x| \rightarrow \infty$ ($\gamma_0 = -{\rm Im}~\omega_0$), rendering the integrals 
(\ref{eq:matelm}) and (\ref{eq:norm}) divergent.
It is therefore critical to handle the asymptotic regions carefully, which,
as we shall see, may be regarded as a way of giving meaning to or regularizing these
formally divergent expressions.

One can deal with the asymptotic regions in several slightly different ways \cite{lpt}, and we 
here give a simple version
that relates most directly to the actual numerical implementation
of the program.  To do so, we return to (\ref{eq:gn}), insert
the integrating factor $\exp [ 2\int dx f_0(x) ]=\phi_0^2(x)$, and integrate 
from $L_{-}$ to $L_{+}$:

---int1
\begin{equation}
\left. \phi_0^2(x) g_n(x) \right|_{L_{-}}^{L_{+}} +
2 \omega_0 \omega_n \int_{L_-}^{L_+} dx\, \phi_0^2(x) \, = \,
\int_{L_{-}}^{L_{+}} dx\, \phi_0^2(x)\, V(x)~.
\label{eq:int1}
\end{equation}
(Had we taken $L_{\pm} = \pm \infty$ instead of finite
values, we would return to the
formal result stated above.)

Now it has to be noticed that $g_n$ contains implicitly the unknown
$n$th order frequency shift $\omega_n$.  To see this, we note that
$g(L_{\pm})$ is to be matched to $f_{\pm}(\omega, L_{\pm})
-f_0(L_{\pm})$; consequently its $n$th-order part is 

---gnmatch
\begin{eqnarray}
\mu^n g_n &=& \left[ f_{\pm}(\omega,L_{\pm}) - f_0(L_{\pm}) \right]_n
= \left[ f_{\pm}(\omega,L_{\pm}) - f_{\pm}(\omega_0,L_{\pm}) \right]_n
\nonumber \\
&=& \left[ f_{\pm}(\omega_0 + \mu \omega_1 + \cdots + \mu^{n-1} 
\omega_{n-1},L_{\pm})
-f_{\pm}(\omega_0,L_{\pm}) \right]_n + \mu^n \omega_n{\partial \over \partial \omega }
f_{\pm}(\omega_0,L_{\pm})
\nonumber \\
&\equiv& \mu^n \left[ \Delta_{\pm n} + \omega_n{\partial \over \partial \omega 
}f_{\pm}(\omega_0,L_{\pm})  
\right]~,
\label{eq:gnmatch}
\end{eqnarray}
where the subscript on the square brackets
indicates taking the $n$th-order part, and
we have separated out the term that depends on $\omega_n$.  
The quantities $\Delta_{\pm n}$ express the difference between two
logarithmic derivatives at the asymptotic points.
Thus, collecting
the terms that do and do not contain $\omega_n$, we obtain from (\ref{eq:int1})
again an expression like (\ref{eq:result}), except that the matrix element and
norm are now given by

---matelm2
\begin{equation}
\langle \phi_0 | V_n | \phi_0 \rangle  =
\int_{L_{-}}^{L_{+}} V_n(x) \phi_0^2(x) dx \,+\, \Delta_{-n}\phi_0^2(L_-)
-\Delta_{+n}\phi_0^2(L_+) 
\label{eq:matelm2}
\end{equation}

---norm2
\begin{equation}
\langle \phi_0 | \phi_0 \rangle =
\int_{L_{-}}^{L_{+}} \phi_0^2(x) dx \,+\, \frac{1}{2\omega_0}\left[ 
\phi_0^2(L_+){\partial \over \partial \omega }f_{+ }(\omega_0,L_{+ })-
\phi_0^2(L_-){\partial \over \partial \omega }f_{- }(\omega_0,L_{-}) \right]~.
\label{eq:norm2}
\end{equation}

Thus, provided we have a way of calculating the values of 
$f_{\pm}(\omega,L_{\pm})$ at the
asymptotic positions $L_{\pm}$, these alternate expressions
(\ref{eq:matelm2}) and (\ref{eq:norm2}) then provide
practical ways of evaluating the matrix element and the norm, without
infinite integrals or divergent expressions.  In other words, the
difficulties alluded to earlier have been eliminated.
The choices of $L_{\pm}$ are arbitrary, and the shifts $\omega_n$ must
be independent of these choices; in fact, it is easy to see that (\ref{eq:matelm2}) and
(\ref{eq:norm2}) must be separately independent of $L_{\pm}$, simply
because $(\ref{eq:norm2})$ relates only to the unperturbed state, whereas
(\ref{eq:matelm2}) depends on the arbitrary perturbation.  This independence
can be verified both analytically by differentiating the expressions with respect
to $L_{\pm}$ and using the Riccati equation, or numerically by evaluating
the expressions for different values of $L_{\pm}$.

---
\section{General Properties}\label{sect:genprop}

The properties of the perturbed QNM's can be discussed at three levels
of generality: (a) properties about open systems in general (in
contrast to conservative systems); (b) properties for any perturbation
of a Schwarzschild black hole; and (c) results for a specific
perturbation.  This Section deals with the first two, and a specific
model of perturbation is presented in \mbox{Sect.\ \ref{sect:shell}}.
Although the LPT is valid to all orders, for the present application
we focus on the first-order shift.

\subsection{Open systems in general}

The result in (\ref{eq:result}) has been written in a way
formally similar to the hermitian case.
The factor $2\omega_0$
occurs because the eigenvalue is $\omega^2$ rather than $\omega$. Since 
the numerator and the denominator in (\ref{eq:result}) are
separately independent of  $L_{\pm}$, 
they can be given physical interpretations as a
generalized matrix element and a generalized norm respectively.

The generalized norm has been introduced previously in a restricted
form \cite{norm} applicable only to cases where the potential has ``no tail";
in that case $\Delta_{\pm } $ can be obtained readily, since 
$f_{\pm }(\omega,L_{\pm}) = \pm i\omega$. 
  In that restricted
form, the perturbation theory for the QNMs of the wave equation has been 
developed and applied \cite{pert}.  For systems such as black holes, the 
``no tail" condition must be removed, as sketched in Sect.~II.
In either case, the generalized norm
has some unusual properties; 
(a) It involves $\phi_0^2$ rather than $|\phi_0|^2$, and is in general 
complex.  
(b) It involves surface terms at $x = L_{\pm}$, though the value of the 
entire expression
is independent of the choice of $L_{\pm}$.  Thus, it is not a 
norm in the strict sense, but rather a useful bilinear map.
In cases where the system parameters can be tuned so that the
leakage of the wavefunction approaches zero (e.g., $V_0(x)$
contains a tall barrier on both sides), then 
the generalized norm reduces to the usual (real and positive-definite)
norm for a NM.  

It is useful to define a function $H(x)$ for each QNM which depends only on the
original unperturbed system: 

---hdef
\begin{equation}
\frac{\delta \omega}{\delta ( \mu V_1(x))} \equiv H(x) 
= \frac{\phi_0(x)^2}{ 2 \omega_0 \langle \phi_0 | \phi_0 \rangle }~. 
\end{equation}
Both the magnitude and the phase of $H(x)$ are
well defined and physically significant.  The magnitude implies that
we can now give a precise meaning to the normalization of a QNM, 
even though the wavefunction diverges at infinity.  The phase of $H(x)$ 
determines the phase of the first-order shift
$\omega_1$ for a real and positive localized perturbation $V_1(x)$. 
The phase is intriguing because it has no counterpart for a hermitian 
system --- in that case, $H(x)$ 
must be real and non-negative. For an open system, $H(x)$ can be negative 
or indeed have any phase, so that a positive perturbation could lead to a 
{\it decrease} of the frequency.

The functions $H(x)$ are then convenient objects for discussing the effect of 
any perturbation on the QNMs of a given system.  We next present some 
properties of $H(x)$ for the Schwarzschild black hole.

---
\subsection{Schwarzschild black holes\label{sect:sch}}

First consider the normalization of these QNMs, made possible by the
introduction of the generalized norm.  Since (\ref{eq:norm})
is independent of $L_{\pm}$, it is particularly convenient if we take
$L_- = L_+ = 0$:

---defA
\begin{equation}
\langle \phi_0 | \phi_0 \rangle =
\frac{1}{2\omega_0} \phi_0(0)^2
\left\{ {\partial \over \partial \omega }f_{+ }(\omega_0,0)-
{\partial \over \partial \omega }f_{- }(\omega_0,0) \right\}
\equiv \phi_0^2(0) K~.
\label{eq:defA}
\end{equation}
The logarithmic derivatives at $x=0$ can be obtained by integrating
the differential equations from $x = \pm \infty$, and are readily
calculated making use of Leaver's solution \cite{leaver} (see Appendix
\ref{app:leaver}).  The parameter $K$ is a convenient way of
expressing the normalization, and Table~\ref{tab:one} lists the values
of $K/(2M)$ for the lowest few QNMs (labeled by $j$) of the
Regge-Wheeler potential for each angular momentum $l$. They increase
as $j$ increases and show clear patterns, e.g., for $l=0$, $s=0$; $K$
is large for odd $j$, while for $l=2$, $s=2$; $K$ is large for even
$j$.  It is interesting to ask in what way the values of $K$ can
characterize the Regge-Wheeler potentials, and to investigate their
meaning beyond the perturbation calculation (noting that $K$ has
nothing to do with the perturbation).



\begin{figure}
\centerline{\epsfig{file=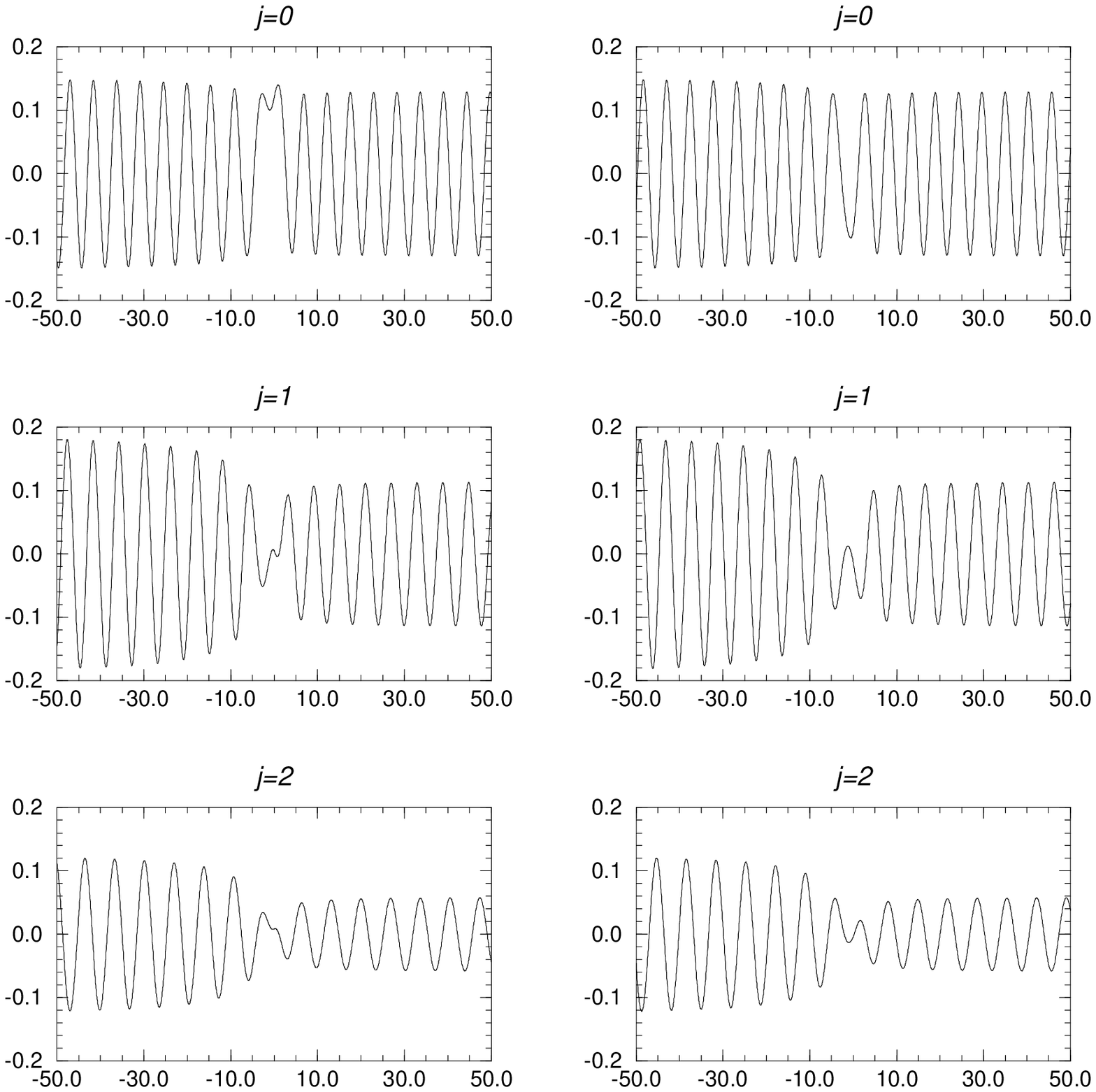,height=10cm,angle=270}}
\centerline{\epsfig{file=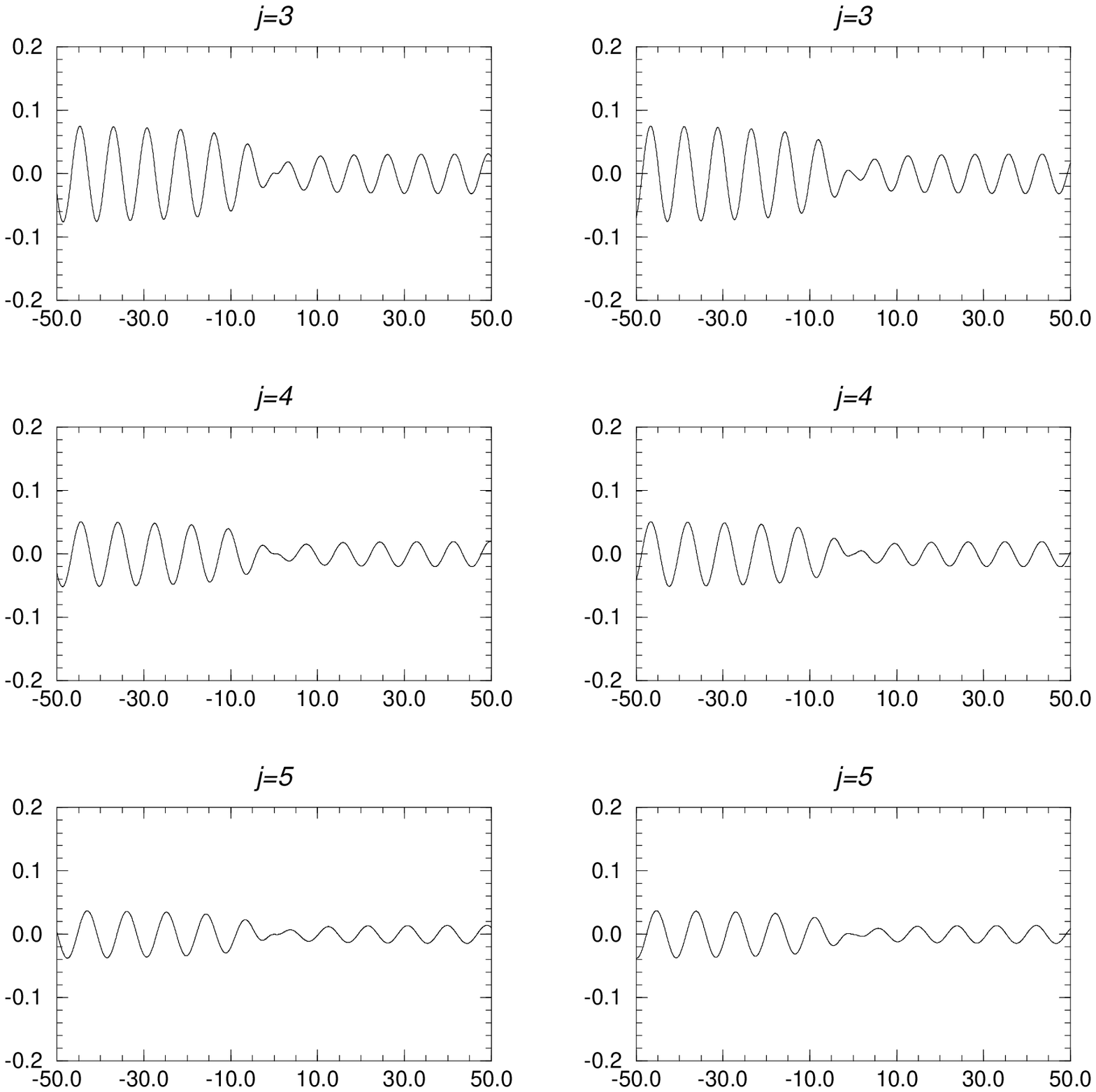,height=10cm,angle=270}}
\vskip 2mm
\caption{Graph of \mbox{Re $H(x)e^{-2\Gamma \protect\sqrt{1+(x/2M)^2}}$}
(left) and 
\mbox{Im $H(x)e^{-2\Gamma \protect\sqrt{1+(x/2M)^2}}$}
(right) vs $x/2M$ for $l=1$, $s=0$, and $j=0,1,\cdots 6$.}
\label{fig:two}
\end{figure}

Next consider the functions $H(x)$.
\mbox{Fig.\ \ref{fig:two}} shows \mbox{Re $H(x)$} 
and \mbox{Im $H(x)$} for $l=1$ scalar waves, versus the 
coordinate $x/(2M)$. 
The diagrams refer to the lowest QNMs (labeled as $j = 0, 1, \cdots , 5$) 
of that angular momentum.  Since the wavefunctions grow exponentially
as $|x| \rightarrow \infty$, we have multiplied $H(x)$ by 
\mbox{$\exp [-2\Gamma \sqrt{1+(x/2M)^2}]$} ($\Gamma =$
\mbox{ $-$Im $2M\omega_0$}) for ease of plotting. 
These figures show an essentially plane wave behavior for large $|x|$,
but the phases are nontrivial. 
We note that for a localized perturbation $\mu V_1(x) = \mu \delta(x-x_1)$,
the first-order shift in frequency $\omega_1$ is given by $H(x_1)$, and 
therefore can be read out directly from the figures.
For example, Table~\ref{tab:two} lists the
magnitude and phases of $H(x)$ for $l=1$, $s=0$ and the lowest few $j$'s
at a fixed value $x=x_1$ ($r = r_1$), for $x_1/2M\ =$ 0, 5, 10 ($r_1/2M\ =$
1.28, 3.93, 8.05); $r_1/2M\ =$ 1.02 and 1.08 ($x_1/2M\ =$ -2.89, -1.45). 
The patterns are different for these values of $x_1$ and not simple, 
demonstrating that a localized perturbation will
push the QNMs along different directions in the complex frequency
plane.



\begin{figure}
\centerline{\epsfig{file=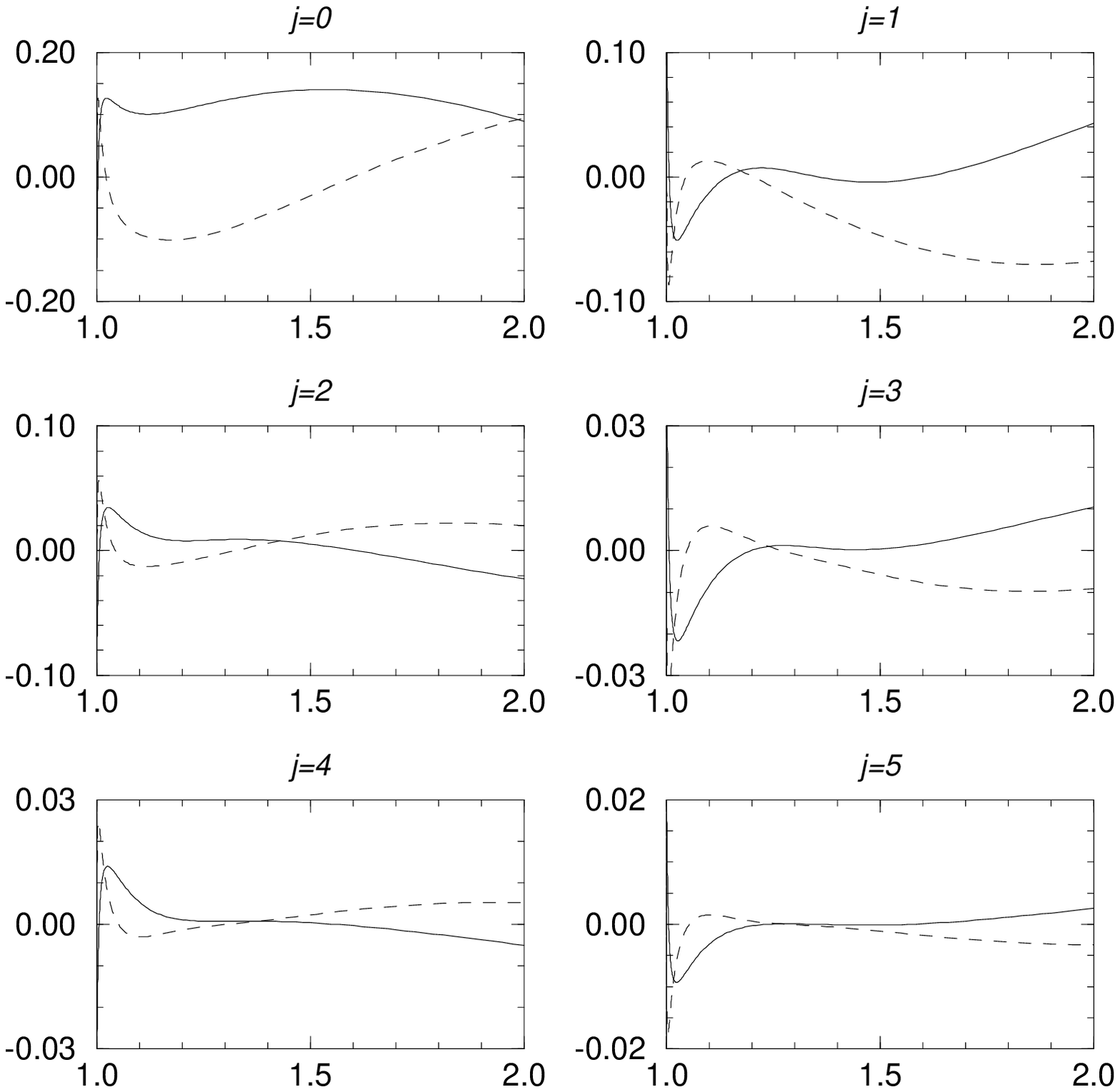,height=15cm}}
\vskip 2mm
\caption{Graph of \mbox{Re $H(x)e^{-2\Gamma \protect\sqrt{1+(x/2M)^2}}$}
(solid line) and \mbox{Im $H(x)e^{-2\Gamma \protect\sqrt{1+(x/2M)^2}}$} 
(dash line) vs $r/2M$.}
\label{fig:three}
\end{figure}

There is a more
complicated structure in $H(x)$ for smaller $x$, best exhibited
if the same data are plotted versus $r/2M$, as in 
\mbox{Fig.\ \ref{fig:three}}. (The region
$r/2M < 2.0$ corresponds to $x/2M<2.0$.)  The many oscillations
for $x<0$ are compressed into a small region near $r/2M=1$ and are not
visible in this plot.  One interesting feature is that both $\mbox{Re $H(x)$}$ 
and $\mbox{Im $H(x)$}$ alternate in sign as $(-1)^j$ near the event horizon. 
The magnitude and phase of $H(x)$ at the sample positions $r=r_1$ where 
$r_1/2M=1.02$ ($x_1/2M \approx -2.89$) and $r_1/2M=1.08$ 
($x_1/2M \approx -1.45$) are also shown in Table~\ref{tab:two}.

The results here described depend only on $H(x)$,
i.e., on the properties of the unperturbed system.
These examples indicate that localized
perturbations can generate a rich pattern of frequency shifts
(in contrast to shifts all of the same phase in the case of 
the NMs of a conservative system).  In turn, this means
that there are much better prospects of learning something
about the perturbing potential from the observed shifts.

Of course, the richness of the pattern could be diluted
if the perturbation is not localized in $x$, but has a spatial
extent $\Delta x$ large compared to the typical wavelength $\lambda$ of 
oscillation of $H(x)$, $\lambda \approx 2\pi / \mbox{Re $\omega_0$} 
\sim$ a few $M$ (see \mbox{Fig.\ \ref{fig:two}} for example). 
In the next Section we evaluate and discuss the perturbation
that arises from the presence of a dust shell, which leads to a perturbation 
potential extending over a range of $x$. 

---

\section{A Model Problem}\label{sect:shell}

\subsection{Description of the model}

We consider a Schwarzschild black hole surrounded by a static
spherical shell of matter at a certain radius outside the hole.  The
equation of state of the matter making up the shell is chosen so that the
shell remains static at the given radius.  This example 
demonstrates that even such a simple model
can lead to intriguing features in the QNM spectrum of the perturbed system.
 
There are two mass parameters in the black hole plus shell system:  
the total ADM mass of the system measured at
infinity $M_o$, and the mass of the black hole as measured by its
horizon surface area $M_a$ (i.e., the surface area of the event
horizon is $16\pi M_a^2$).  In the limit of $M_a \rightarrow M_o$, we
return to the unperturbed case of a bare black hole.  Thus the parameter $\mu \equiv 
(M_o-M_a)/M_a$ is a measure of the perturbation.  The
perturbation further depends on the circumferential radius $r=r_s$
where the shell is placed.  We shall study the shifts as a function of
$\mu$ and $r_s$.

A static and spherically symmetric metric can be written as 
---metric
\begin{equation}
  ds^2=-A(r) dt^2+B(r) dr^2+r^2(d\theta^2+\sin^2\theta^2 
d\varphi^2)~.
\label{appe:metric}
\end{equation}
As usual, we define the mass function $m(r)$ such that 
\begin{equation}
  g_{rr}=B(r) =\left[1-\frac{2m(r)}{r}\right]^{-1}~.
\end{equation}
The functions $m(r)$ and $A(r)$ satisfy the equations 
\begin{eqnarray}
  \frac{dm}{dr} &=& 0 \qquad \qquad , \qquad \qquad r \neq r_s \\
 & & \cr
  {1 \over 2A} \frac{dA}{dr} &=& \frac{m}{r(r-2m)}~.  
\end{eqnarray}
Hence we have 
\begin{equation}
  m(r)=\left \{ \begin{array}{ll}
	M_o \qquad & r>r_s \\
	M_a \qquad & r<r_s 
\end{array} \right.
\end{equation}
\begin{equation}
  A(r)=\left \{ \begin{array}{ll}
	1-\bfrac{2M_o}{r} \qquad & r \geq r_s \\
	& \\
	\alpha \left( 1-\bfrac{2M_a}{r}\right) \qquad & r<r_s
\end{array} \right.~,
\end{equation}
where 
\begin{equation}
  \alpha=\frac{1-2M_o/r_s}{1-2M_a/r_s}~. 
\end{equation}
The metric (\ref{appe:metric}) becomes 
---metric2
\begin{equation}
ds^2=\left \{ \begin{array}{ll}
-\left(1-\bfrac{2M_o}{r}\right)dt^2+\left(1-\bfrac{2M_o}{r}\right)^{-1}dr^2 
+r^2(d\theta^2+\sin^2 \theta d\varphi^2) \ \ \ \ \ & r>r_s \\
 & \\
 -\alpha \left(1-\bfrac{2M_a}{r}\right)dt^2
+\left(1-\bfrac{2M_a}{r}\right)^{-1}
dr^2+r^2(d\theta^2+\sin^2 \theta d\varphi^2) \ \ \ \ \ & r<r_s
	\end{array} \right.~.
\label{appe:metric2}
\end{equation}
The physical meanings of the parameters $r$, $M_o$ and 
$M_a$ are now clear from (\ref{appe:metric2}). We see explicitly
that in the limit $M_o \rightarrow M_a$, the value of $\alpha$ 
reduces to unity, and (\ref{appe:metric2}) reduces to the familiar 
Schwarzschild metric. 

Next we consider waves propagating on this perturbed Schwarzschild background.
In this paper for simplicity we focus on the case of
scalar wave ($s=0$ in (\ref{eq:schw})), described by 
\begin{equation}
\partial_{\mu} \partial^{\mu} \psi = \sqrt{-g} \partial_{\mu} \left( \sqrt{-g}
g^{\mu \nu} \partial_{\nu} \psi \right) = 0~. 
\end{equation}
To obtain the Klein-Gordon equation suitable for the perturbation analysis,
we separate $\psi$ into the radial and angular parts: 
\begin{equation}
  \psi({\bf r},t)=\sum_{l=0}^{\infty} \sum_{m=-l}^l R_l(r,t) 
Y_{lm}(\theta,\varphi)~. 
\end{equation}
The radial function satisfies 
\begin{equation}
  \frac{A}{B}R_l''+\left(\frac{2A}{rB}+\frac{A'}{2B}-\frac{AB'}{2B^2}
\right) R_l'-\frac{Al(l+1)}{r^2}R_l=\partial_t^2 R_l~,
\end{equation}
where $'\equiv d/dr$. 
Next introduce the tortoise coordinate 
\begin{equation}
  x=\int^r \sqrt{\frac{B}{A}}\, dr
\label{appe:rstar}
\end{equation}
to push the event horizon ($r=2M_a$) to $-\infty$. We leave the
zero point of $x$ to be specified later. The equation for 
$R_l$ becomes 
\begin{equation}
  \partial_x^2 R_l + \frac{2}{r}\sqrt{\frac{A}{B}}\, \partial_x R_l
-\frac{Al(l+1)}{r^2}R_l=\partial_t^2 R_l~.
\end{equation}
To cast the equation into the standard Klein-Gordon form, we introduce
a function $\Phi(x,t)$ (hereafter the label $l$ will be suppressed) such that 
\begin{equation}
  R(x,t)=h(x)\Phi(x,t)~,
\end{equation}
with 
\begin{equation}
  \partial_x h+\frac{1}{r}\sqrt{\frac{A}{B}}h=0~. 
\end{equation}
This function $\Phi(x,t)$ then satisfies the Klein-Gordon equation 
(\ref{eq:kg}) with the effective potential
\begin{equation}
  V(r)=\frac{Al(l+1)}{r^2}+\frac{A'}{2Br}-\frac{AB'}{2rB^2}~.
\end{equation}
Substituting in the expressions for $A$ and $B$, we get 
\begin{equation}
  V(r)=\left\{ \begin{array}{ll} 
   \left(1-\bfrac{2M_o}{r}\right)\left[\bfrac{l(l+1)}{r^2}
+\bfrac{2M_o}{r^3}\right]\equiv 
V_{SC}(M_o,r) \qquad & r>r_s \\
   \kappa_r \, \delta (r - r_s) \qquad & r=r_s \\
    \alpha \left(1-\bfrac{2M_a}{r}\right)
\left[\bfrac{l(l+1)}{r^2}+\bfrac{2M_a}{r^3}\right]=\alpha V_{SC}(M_a,r) 
\qquad & r<r_s 
  \end{array}\right.~,
\label{eq:fullpot}
\end{equation}
where $V_{SC}(M,r)$ is the standard unperturbed potential for a Schwarzschild 
black hole with mass $M$ (i.e., equation~(\ref{eq:schw}) with $s=0$). 
The strength of the $\delta$-function at the position of the shell $r_s$ 
is given by 
\begin{equation}
  \kappa_r=\lim_{\epsilon \rightarrow 0^+} 
\int_{r_s-\epsilon}^{r_s+\epsilon} V(r)\, dr
=-\left(1-\bfrac{2M_o}{r_s}\right)\bfrac{\mu M_a}{r_s^2}~.
\end{equation}

The remaining degree of freedom in setting up the perturbed problem is
the zero-point of the tortoise coordinate $x$.  We choose it so that
the effective potential is the standard unperturbed form {\it outside}
the shell, in terms of the ADM mass. The zero point is then determined by the 
requirement of continuity of $x$.
\begin{equation}
  x=\left\{ \begin{array}{ll}
  r+2M_o\ln(r/2M_o-1) ~~~~~~~~~ & r \geq r_s \\
  \bfrac{1}{\sqrt{\alpha}}
\left[r+2M_a\ln(r/2M_a-1)\right]+x_c ~~~~~ & r<r_s 
	\end{array} \right.~,
\end{equation}
where $x_c$ is the constant which makes $x$ continuous. 

Thus we arrive at the perturbation problem (\ref{eq:kg1}) as discussed in 
the sections above with $x$ being the tortoise coordinate, 
\begin{eqnarray}
V_0 (x) &=& V_{SC}(M_o,x) \\
 & & \cr
\mu V_1(x)  &=& \left\{ \begin{array}{ll} 
      0  ~~~~~~~~~~~~~~ & \mbox{for }  x > x_s \\
      \kappa \delta (x - x_s) + \alpha V_{SC}(M_a, r)-V_{SC}(M_o, r)
~~~~~~~~~~~~~~ & {\rm otherwise}
	\end{array}\right.~,
\label{eq:shellv1}
\end{eqnarray}
where 
\begin{equation}
  x_s=r_s+2M_o\ln \left(\frac{r_s}{2M_o}-1\right) 
\end{equation}
and 
\begin{equation}
  \kappa=\lim_{\epsilon \rightarrow 0^+} \int_{x_s-\epsilon}^{x_s+\epsilon} 
V(x)\, dx=-\bfrac{2\mu M_a}{r_s^2 (1+1/\sqrt{\alpha})}~.
\label{eq:lambda}
\end{equation}
This perturbation consists of two parts: a $\delta$-function at the position
of the shell, plus a contribution inside the shell extending all the way
to the horizon ($x \rightarrow -\infty$ or $r \rightarrow 2M_a$). 
There is no perturbation outside the shell. 

We should make one further comment concerning the setup of the problem. 
As the equation of state of the matter shell is chosen to make the shell 
static, one need to ensure that the equation of
state satisfies the energy conditions.  It is straightforward to show that
the dominant 
energy condition is the first to be violated if the shell is placed too close
to the horizon, as shown in Appendix \ref{app:endom}. Thus, the discussion 
will be limited to the perturbations that do not violate the dominant 
energy condition. 

\subsection{Numerical solution}
The full potential $V$ can be cast into a standard Regge-Wheeler potential
for both $x < x_s$ and $x > x_s$, and hence exact QNMs can be obtained
by numerically evaluating the Leaver's solutions of the Regge-Wheeler 
equation (see \cite{leaver} and Appendix~\ref{app:leaver}). The numerical
solution will be used to examine the accuracy and validity of the perturbative result.
Details of the numerical scheme are sketched as follows.

Firstly, as there is a $\delta$-function in the effective
potential $V(x)$, the logarithmic derivatives at $x=x_s$ are related by 
\begin{equation}
  \lim_{\epsilon \rightarrow 0^+} \left[ \frac{\phi'(\omega,x_s+\epsilon)}
{\phi(\omega,x_s+\epsilon)}-\frac{\phi'(\omega,x_s-\epsilon)}
{\phi(\omega,x_s-\epsilon)} \right]=\kappa \ \ .
\label{qnmcond1}
\end{equation}

Secondly, outside the shell, the effective potential is the Regge-Wheeler potential 
with mass parameter $M_o$ and the QNM wave function satisfies the outgoing wave
boundary condition at the spatial infinity $x \rightarrow \infty$. Hence,
\begin{equation}
  \phi (\omega,x)=\phi_{+}(2M_o\omega,x/2M_o) \ \ ,
\end{equation}
where $\phi_{+}(\tilde{\omega},\tilde{x})$ is the outgoing wave solution of 
the scaled Klein-Gordon equation 
\begin{equation}
  \left[ \frac{d^2}{d\tilde{x}^2}+\tilde{\omega}^2-V(\tilde{r}) \right] 
\phi(\tilde{\omega},\tilde{x})=0
\label{normkgw}
\end{equation}
with 
\begin{equation}
  V(\tilde{r}) = \left(1-\frac{1}{\tilde{r}}\right)\left[\frac{l(l+1)}
{\tilde{r}^2}+\frac{1}{\tilde{r}^3}\right] 
\end{equation}
and
\begin{equation}
	\tilde{x} = \tilde{r}+\ln(\tilde{r}-1) \ \ .
\end{equation}
Note that the quantities with tilde are dimensionless. 

Thirdly, inside the shell, it is also easy to show that the wave function
$\phi$ satisfies
\begin{equation}
  \left[ \frac{d^2}{d\bar{x}^2}+\left( \frac{\omega}{\sqrt{\alpha}} 
\right)^2 - V_{SC}(M_a,r) \right] \phi =0 \ \ \ \ \ x <x_s \ \ ,
\end{equation}
and 
\begin{equation} 
  \bar{x}=r+2M_a \ln \left( \frac{r}{2M_a}-1 \right)=\sqrt{\alpha}
(x-x_c) \ \ .
\end{equation}
Hence 
\begin{equation}
  \phi(\omega,x)=\phi_{-}(2M_a \omega/\sqrt{\alpha},\bar{x}/2M_a) \ \ .
\end{equation}
Here $\phi_{-}(\tilde{\omega},\tilde{x})$ is the ingoing wave solution 
of~(\ref{normkgw}) at $\tilde{x} \rightarrow -\infty$.

Therefore, by connecting the two logarithmic derivatives at
$x=x_s$ with~(\ref{qnmcond1}), the QNM condition can be written as 
\begin{equation}
\left. \frac{ \phi'_{+}(2M_o\omega,\tilde{x})}{\phi_{+}(2M_o\omega,\tilde{x})} 
\right|_{\tilde{x}=x_s/2M_o}
- \left. \frac{p \phi'_{-}(2M_o\omega/p,\tilde{x})}
{\phi_{-}(2M_o\omega/p,\tilde{x})} 
\right|_{\tilde{x}=\bar{x}_s/2M_a} = 2M_o \kappa \ \ ,
\label{qnmcond2}
\end{equation}
where $p=\sqrt{\alpha}M_o/M_a$, 
\begin{equation} 
  \bar{x}_s=r_s+2M_a \ln \left( \frac{r_s}{2M_a}-1 \right)=\sqrt{\alpha}
(x_s-x_c) \ \ ,
\end{equation}
and the prime represents the derivative with respect to $\tilde{x}$. The 
functions $\phi_{+}$ and $\phi_{-}$
can be computed by the Leaver's solutions of the Regge-Wheeler 
equation (see \cite{leaver} and Appendix~\ref{app:leaver}). 
Therefore, exact QNMs can be obtained by solving the nonlinear 
equation~(\ref{qnmcond2}) by standard root-searching methods.  

\subsection{Dependence on $\mu$ and convergence}
Given the perturbation function $\mu V_1$ in (\ref{eq:shellv1}), the shifts 
$\omega_1$ and $\omega_2$ can be evaluated by 
(\ref{eq:result}). The detailed treatments will be given in Appendix 
\ref{app:pert}. 


\mbox{Fig.\ \ref{fig:four}} shows the 0th, 1st and 2nd order perturbation 
results for $l=1$ scalar waves 
together with the exact numerical results for a black hole plus shell system.
The parameters are $\mu = 0.02$, 
$r_s=2.52 M_a$ ($x_s \approx 0$).  This figure shows that
the perturbation formalism does give the correct shifts for small
$\mu$. 

\begin{figure}
\centerline{\epsfig{file=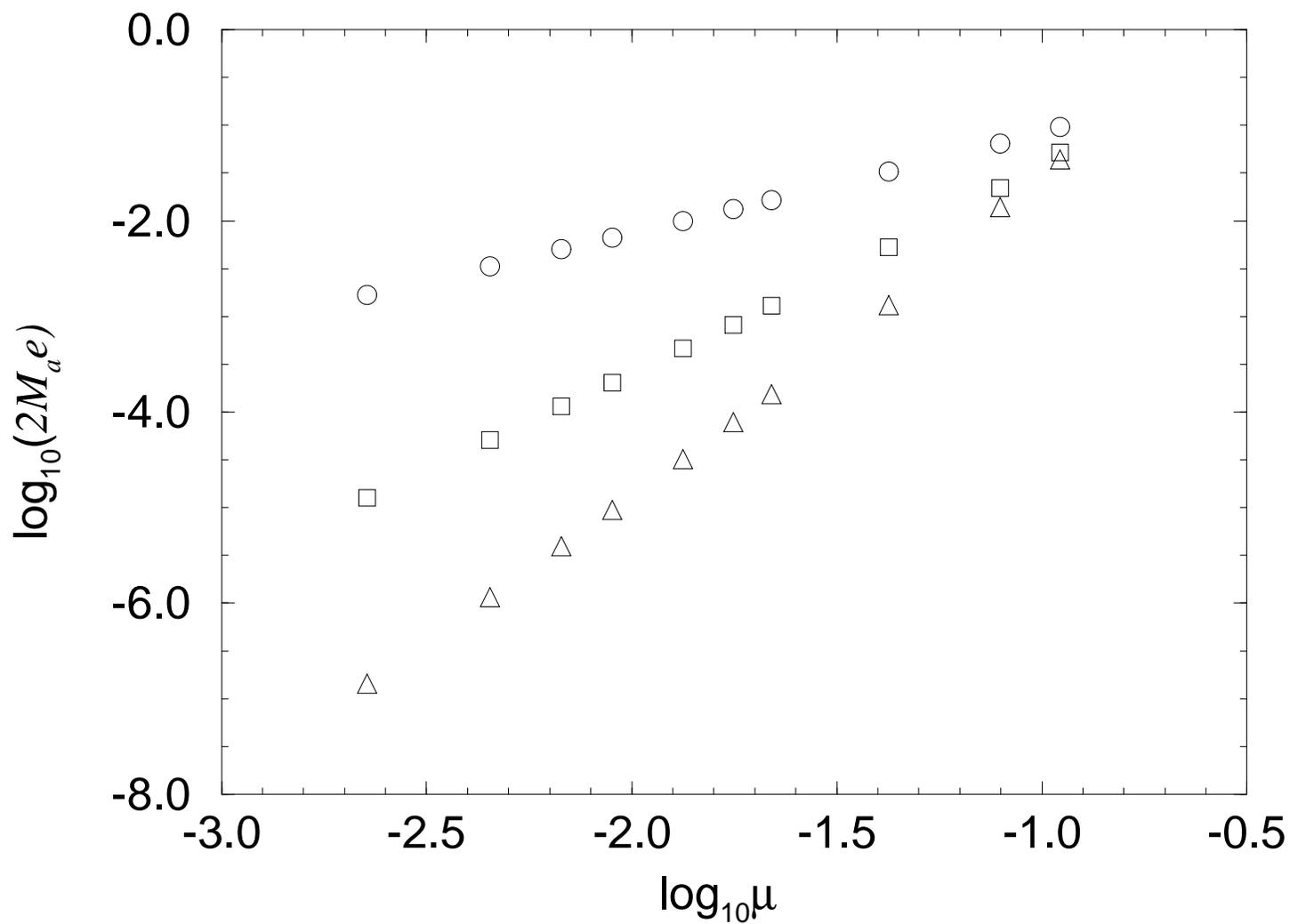,height=15cm}}
\vskip 2mm
\caption{The magnitude of the error in the frequencies of
the 0th, 1st and 2nd order perturbation for $l=1$, $s=0$, $j=1$, and
$r_s=2.52 M_a$.}
\label{fig:five}
\end{figure}

To see the convergence more clearly, we
plot in \mbox{Fig.\ \ref{fig:five}} the magnitude of the error in the 
frequencies in the 0th, 1st and 2nd order results versus $\mu$.  
The plot shows the case of
$l=1$, $s=0$, $j=1$ (first excited state), with the shell located at
$r_s=2.52M_a$. The error of the $n$th-order result goes as 
$\mu^{n+1}$, as it should.


\subsection{Dependence on shell position}

\begin{figure}
\centerline{\epsfig{file=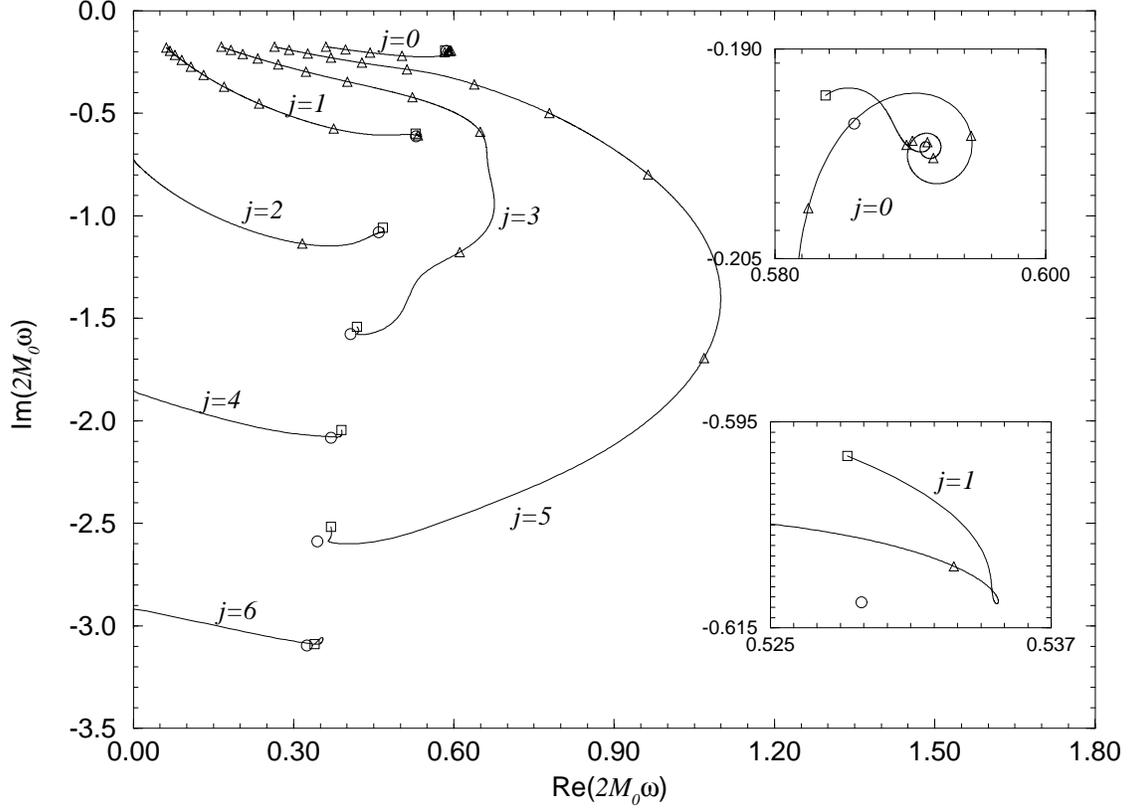,height=15cm,angle=270}}
\vskip 2mm
\caption{The trajectory of the lowest QNMs of $l=1$ scalar waves for
$\mu=0.01$ and $r_s/M_a$ varies from $2.26$ to $60$ based on exact 
numerical calculation.
The circles are QNMs of a bare Schwarzschild black hole with mass $M_o$; the
squares are the QNMs for $r_s=2.26 M_a$ (The dominant energy condition is
violated when $r_s<2.26 M_a$); the triangles show the positions
of QNMs at $r_s/M_a$ from 6 to 60 in intervals of 6.
The upper and lower inset
show the regions near $j=0$ and $j=1$ in details respectively.
We stopped the calculation when the QNMs approach the imaginary $\omega$ axis
because it is difficult to compute it accurately in that region. Thus, some
triangles appear ``missing'' for some higher damping QNMs because they have
already moved to the imaginary $\omega$ axis.}
\label{fig:six}
\end{figure}

We next study the dependence on the parameters of the shell. 
\mbox{Fig.\ \ref{fig:six}} shows the trajectories of the lowest damping 
QNMs ($j=0,1,\cdots 6$) as 
the position of the shell 
moves away from the event horizon. The plot shows the case of $l=1$ 
scalar waves with $\mu=0.01$ based on numerical calculation. The regions 
near the $j=0$ and $j=1$ modes are shown 
in greater detail in the insets.  The calculation is terminated when the QNM 
is near the imaginary $\omega$ axis, since it is difficult to perform 
the numerical calculation with sufficient 
accuracy in that region. It is seen that there are interesting changes 
in signs with increasing $j$. Some QNMs move toward the imaginary 
$\omega$ axis, and the higher damping QNMs either move 
toward the imaginary $\omega$ axis or move upward toward the 
origin. It is also seen that when the shell 
is placed far away from the event horizon, the QNMs move away from their 
unperturbed positions, with the higher damping modes moving with higher speed. 
This behavior can be understood from the perturbation formula 
(\ref{eq:result}). When $x_s/2M_a \gg 1$, we have, according to 
(\ref{eq:shellv1}) and (\ref{eq:result}), 
---asyom 
\begin{equation}
\omega_1 \sim e^{2i\omega_0 x_s}/x_s^2~~~~~\mbox{for~~~~}x_s/2M_a \gg 1~.
\label{eq:asyom}
\end{equation}
Hence QNMs move more rapidly away from the unperturbed positions when the 
shell is far away from the event horizon ($x_s/2M_a \gg 1$) and the higher 
damping modes (\mbox{$-$Im $2M_a \omega_0 \gg 1$}) move with higher speed. 

\begin{figure}
\centerline{\epsfig{file=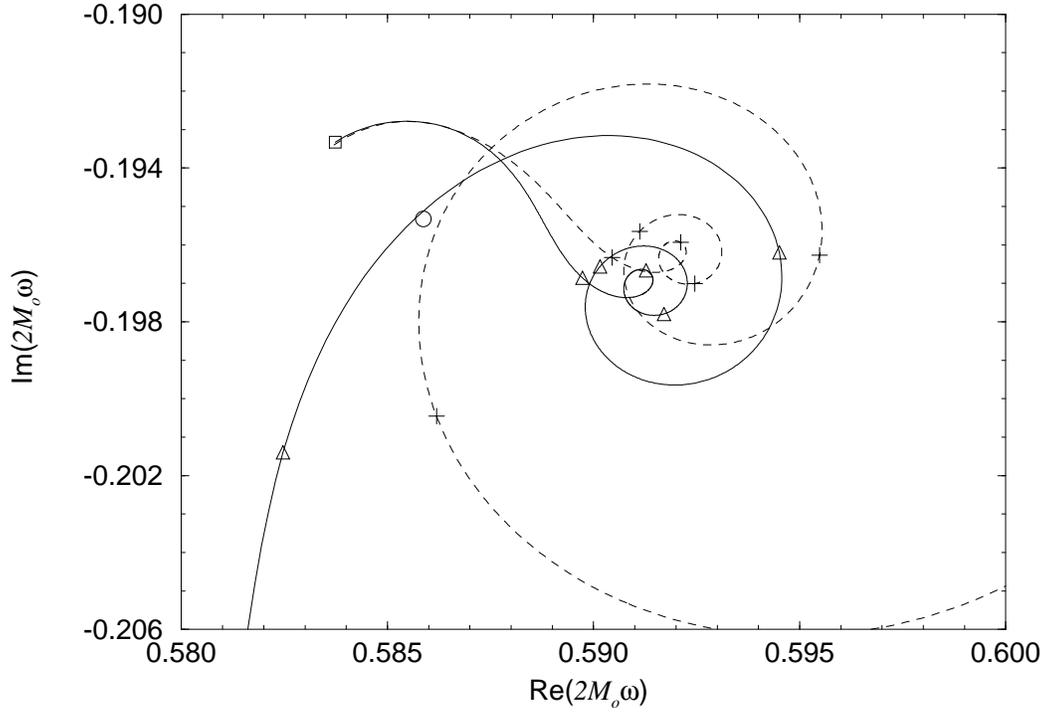,height=14cm,angle=270}}
\vskip 2mm
\caption{Trajectory of $j=0$ mode based on exact (solid line) and first order
(dash line) calculation. The zeroth order result is indicated by a circle and 
the square represents the mode for $r_s=2.26M_a$. 
The triangles show the mode for the positions of the shell at $r_s/M_a$ from 
6 to 36 in intervals of 6 for exact result, and the plus symbols show the 
first order result for the shell at the corresponding positions.}
\label{fig:seven}
\end{figure}

\begin{figure}
\centerline{\epsfig{file=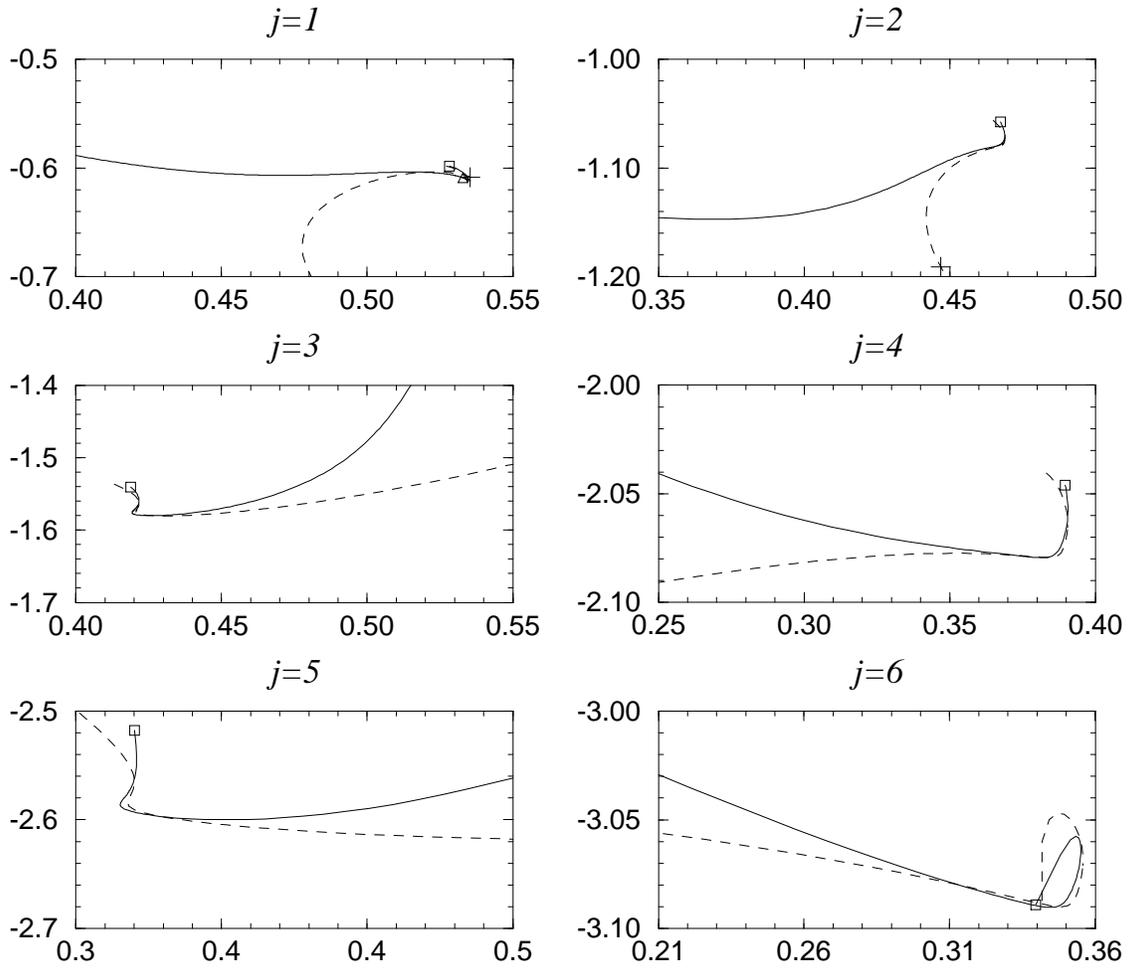,height=15cm,angle=270}}
\vskip 2mm
\caption{Same as \protect\mbox{Fig.\ \protect\ref{fig:seven}} but for 
$j=1,2,\cdots 6$. For these $j$, the perturbation breaks down before a spirl 
forms.}
\label{fig:eight}
\end{figure}

\mbox{Fig.\ \ref{fig:seven}} shows the first-order perturbation result for 
the $j=0$ mode for different values of $x_s$. At large $x_s$, the trajectory 
shows a spiral structure, which can be explained from the first-order 
perturbation formula. According to (\ref{eq:asyom}), 
\begin{equation}
\frac{d\omega}{dx_s} \approx \frac{d\omega_1}{dx_s} \sim 
\frac{e^{2i\omega_0 x_s}}{x_s^2} \qquad \mbox{for }x_s \gg 1~.
\end{equation} 
The exponential factor $e^{2i\omega_0 x_s}$ leads to the spiral 
structure. For larger $j$ modes, there is no such spiral 
structure (\mbox{Fig.\ \ref{fig:six}}) because 
the higher-order corrections become large before the spiral appears, as shown 
in \mbox{Fig.\ \ref{fig:eight}}. 

---
\section{Conclusion}\label{sect:con}

We have applied the LPT for the QNMs of open systems to the study of
gravitational waves propagating away from black holes.  QNM
gravitational wave signals from black holes will be detected soon, and
many black holes are expected to be perturbed by their astrophysical
environment, e.g., by an accretion disk. The study in this paper on
scalar waves represents the first step in the study of the QNM spectra
of waves propagating in a dirty black hole background, perturbed by
the astrophysical environment.  Its further development can be of
interest to gravitational wave astronomy, among other applications.

Although the QNMs of any 
system can in principle be obtained through brute force numerical
integration, it is nevertheless highly desirable to have a convenient
set of perturbation formulae that one can understand the system with.
We note that the Rayleigh-Schr\"{o}dinger perturbation theory for
NMs is tremendously valuable, even though the NM systems are more easily 
handled numerically than QNM systems.

We have shown in a simple example that a perturbed black hole spectrum can
have interesting features, and demonstrated how these features can
be understood with the perturbation formula. In summary, we raise the 
importance of studying QNMs of dirty black
holes, and show how it can be done in a perturbative formulation.

\acknowledgments

This work is supported in part by the Hong Kong Research Grants
Council grant 452/95P and the US NSF grant PHY 96-00507 and NASA grant
NCCS5-153.  WMS also thanks the Institute of Mathematical Science of
The Chinese University of Hong Kong for its support.

\appendix

\section{Dominant Energy Condition}
\label{app:endom}

The shell cannot be placed too near to the black hole; otherwise, the dominant
energy condition 
\begin{equation}
  |{S_{\hat{t}}}^{\hat{t}}|>|{S_{\hat{q}}}^{\hat{q}}| \qquad \mbox{(no sum on 
$q$)}
\label{engcond}
\end{equation}
will be violated, where $\hat{q}$ denotes any unit specelike vector, and 
\begin{equation}
  {S_{\hat{\mu}}}^{\hat{\nu}}=\lim_{\epsilon \rightarrow 0^+} 
\int_{r_s-\epsilon}^{r_s+\epsilon} {T_{\hat{\mu}}}^{\hat{\nu}}\, d\hat{r}~.
\end{equation}

It is straightforward to compute all the
non-trivial components of the Einstein tensor for the metric 
(\ref{appe:metric}): 
\begin{eqnarray}
  {G_r}^r &=& \frac{A'}{rAB}+\frac{1}{r^2B}-\frac{1}{r^2} \\
 & & \cr
  {G_{\theta}}^{\theta}={G_{\varphi}}^{\varphi} &=&\frac{A'}{2rAB}
-\frac{{A'}^2}{4A^2B}-\frac{B'}{2rB^2}-\frac{A'B'}{4AB^2}+\frac{A''}{2AB} \\
 & & \cr
  {G_{t}}^t &=& \frac{1}{r^2B}-\frac{1}{r^2}-\frac{B'}{rB^2}~,
\end{eqnarray}
where $'\equiv d/dr$. 
The component ${G_{\mu}}^{\nu}$ is non-zero only if it contains terms
involving the second derivative of $A$ or the first derivative of $B$. Hence
\begin{eqnarray}
  {S_{\hat{t}}}^{\hat{t}}={S_{t}}^t &=& \frac{1}{8\pi}\,
\lim_{\epsilon \rightarrow 0^+} \int_{r_s-\epsilon}^{r_s+\epsilon}
\frac{-B'}{rB^{3/2}}\, dr \\
 & & \cr
 &=& -\frac{M_o-M_a}{2\pi r_s^2}\left(\sqrt{1-\frac{2M_o}{r_s}}
+\sqrt{1-\frac{2M_a}{r_s}}\right)^{-1}
\label{stt}
\end{eqnarray}
using the Einstein equations \mbox{${G_{\mu}}^{\nu}=8\pi {T_{\mu}}^{\nu}$}. 
Similarly, we have
\begin{eqnarray}
 {S_{\hat{\theta}}}^{\hat{\theta}}={S_{\hat{\phi}}}^{\hat{\phi}} &=&
\frac{1}{8\pi r_s}\left(\frac{1-M_o/r_s}{\sqrt{1-2M_o/r_s}}
-\frac{1-M_a/r_s}{\sqrt{1-2M_a/r_s}}\right)~.
\label{sff}
\end{eqnarray}

(\ref{engcond}) together with (\ref{stt}) and (\ref{sff}) imply a lower 
bound on $r_s$ for a given $\mu$. For $\mu=0.01$, it is found that the 
minimum $r_s$ is $2.26\, M_a$.

\section{Leaver's Solutions}
\label{app:leaver}
 
In the analytic study of the Regge-Wheeler equation, it is
convenient to rewrite the Regge-Wheeler equation in terms of the 
coordinate $r$:
\begin{equation}
  r(r-1)\phi_{,rr}+\phi_{,r}-\left[ \frac{\rho^2 r^3}{r-1}+l(l+1)+
\frac{\xi}{r} \right] \phi=0~,
\label{rwr}
\end{equation}
where $\rho=-i\omega$, $\xi=1-s^2$. For simplicity, we have set $2M=1$. The 
equation (\ref{rwr}) can be transformed to the generalized spheroidal wave 
equation \cite{leaver}. For an arbitrary frequency $\omega$, 
we define two solutions $\phi_{\pm}(\omega,r)$ 
which satisfy the boundary conditions  
\begin{equation}
  \phi_{\pm} \propto  e^{\pm i\omega x} \qquad \mbox{for }x \rightarrow 
\pm \infty  
\end{equation}
The analytic solutions of these two functions are given by \cite{leaver} 
\begin{eqnarray}
  \phi_-(\omega,r) &=& (r-1)^{\rho}r^{-2\rho}e^{-\rho r} 
\sum_{n=0}^{\infty} a_n\left( \frac{r-1}{r} \right)^n \label{lein} \\ 
  \phi_+(\omega,r) &=& r^{1+s}(r-1)^{\rho}e^{-\rho r} \sum_{n=0}^{\infty} 
a_n (2\rho+1)_n\, U(s+1+2\rho+n,2s+1,2\rho r)~,
\label{leout}
\end{eqnarray}
where 
\begin{equation}
  (2\rho+1)_n \equiv \frac{\Gamma(2\rho+1+n)}{\Gamma(2\rho+1)}
\end{equation}
is the Pochhammer's symbol, $U$ is the irregular confluent hypergeometric 
function \cite{handbook}, and $a_n$ is determined by the following three-term 
recursion relation: 
\begin{equation}
  \alpha_n a_{n+1}+\beta_n a_n+\gamma_n a_{n-1}=0 \ \ \ \ \ n=1,2,\cdots~,
\label{recschan}
\end{equation}
with
\begin{equation}
  a_n=0 \ \ \ \ \ \mbox{for } n<0~. 
\label{schintan}
\end{equation}
The value of $a_0$ is arbitrary and is related to the normalization of
$\phi_{\pm}$. The quantities $\alpha_n$, $\beta_n$ and $\gamma_n$ are defined 
to be 
\begin{mathletters}
\begin{eqnarray}
  \alpha_n &=& (n+1)(n+2\rho+1) \\
  \beta_n &=& -\left[ 2n^2+(8\rho+2)n+8\rho^2+4\rho+l(l+1)+\xi
\right] \\
  \gamma_n &=& n^2+4\rho n +4\rho^2+\xi -1~. 
\end{eqnarray}
\label{appeq:coefs}
\end{mathletters}

In most cases, we only need the logarithmic derivatives 
\begin{equation}
  f_{\pm}(\omega,x)=\frac{1}{\phi_{\pm}}\frac{d\phi_{\pm}}{dx}~. 
\end{equation}
While $f_-(\omega,x)$ can be easily obtained from (\ref{lein}), the 
calculation of $f_+(\omega,x)$ is less straightforward as it involves 
a sum of irregular confluent hypergeometric functions, which are notoriously 
different to evaluate \cite{leaver}. In the following, we develop a 
numerical algorithm to calculate $f_+(\omega,x)$. 

We define two sequences 
\begin{eqnarray}
  T_n^{(-)} &=& (2\rho+1)_n \, U(s+1+2\rho+n,2s+1,2\rho r) \label{tn-} \\ 
  T_n^{(+)} &=& \frac{\Gamma(2\rho+1+n)}{\Gamma(2\rho+n-s+1)}
M(s+1+2\rho+n,2s+1,2\rho r)
\label{tn+}
\end{eqnarray}
where $M$ is the regular confluent hypergeometric function. Hence we have  
\begin{equation}
  \phi_+(\omega,r)=r^{1+s}(r-1)^{\rho}e^{-\rho r}
h(\omega,r)~,
\end{equation}
where
\begin{equation}
  h(\omega,r)=\sum_{n=0}^{\infty} a_n T_n^{(-)}~. 
\label{serh}
\end{equation}
$ f_+(\omega,x)$ can then be expressed in terms of $h$ as follows:
\begin{equation}
  f_+(\omega,x)= \left(1-\frac{1}{r}\right) 
\frac{1}{\phi_+}\frac{d\phi_+}{dr}=\left(1-\frac{1}{r}\right) \left[ 
\frac{1+s}{r}+\frac{\rho}{r-1}-\rho+\frac{1}{h}\frac{dh}{dr} \right]~. 
\label{logdphi}
\end{equation}

It can be shown, from the recursion formulae of the confluent 
hypergeometric functions \cite{handbook,temme}, that 
\begin{eqnarray}
 & & (2\rho+n)(2\rho+n+1) T_{n-1}^{(\pm)}-(2\rho+n+1)(1+4\rho+2n+2\rho r)
T_n^{(\pm)} \cr
 & & + (s+1+2\rho+n)(2\rho+n-s+1)T_{n+1}^{(\pm)}=0 \label{recT}
\end{eqnarray}
and 
\begin{equation}
  \frac{dT_n^{(-)}}{dr}=\frac{s+1+2\rho+n}{r}\left[ \frac{2\rho+n+1-s}
{2\rho+n+1}T_{n+1}^{(-)}-T_n^{(-)} \right]~.
\end{equation} 
For large $n$, $T_n^{(\pm)}$ approaches the asymptotic expressions \cite{temme}
\begin{eqnarray}
T_n^{(-)} &\approx & \frac{\sqrt{\pi}\, \Gamma(2\rho+1+n)}{\Gamma(2\rho+1) \,
\Gamma(2\rho+1+n+s)}\left( \frac{n}{2\rho r} \right)^s 
(2n\rho r)^{-1/4} e^{\rho r}\exp \left[ -2 (2n\rho r)^{1/2} \right]
\label{asyT-} \\
 & & \cr
T^{(+)}_n &\approx & \frac{\Gamma(2s+1) \, \Gamma(2\rho+1+n)}{2\sqrt{\pi} \,
\Gamma(2\rho+1+s+n)}\left( \frac{n}{2\rho r} \right)^s
(2n\rho r)^{-1/4}e^{\rho r} \exp \left[ +2 (2n\rho r)^{1/2} \right]~. 
\label{asyT+}
\end{eqnarray}
The general solution of the difference equation (\ref{recT}) is a linear 
combination of $T_n^{(\pm)}$. It follows from (\ref{asyT-}) and (\ref{asyT+}) 
that $T_n^{(+)}$ increases with $n$, while $T_n^{(-)}$ decreases with $n$. 
Hence any linear combination of $T_n^{(\pm)}$ will eventually be dominated by 
the term containing $T_n^{(+)}$ as $n \rightarrow \infty$. The computation of 
$T_{n+1}^{(-)}$ from $T_n^{(-)}$ and $T_{n-1}^{(-)}$ is therefore unstable. 
The recursion relation (\ref{recT}) must be used in the reverse direction. 
In practice, we choose an arbitrary value of $T_N^{(-)}$ ($N \gg 1$) and set 
\begin{equation}
  T_{N-1}^{(-)}=\frac{T_{N-1}^{(-)}}{T_N^{(-)}}\times T_N^{(-)} \approx
T_N^{(-)}\, \frac{2\rho+N+s}{2\rho+N}\left(\frac{N-1}{N}\right)^{s-1/4}
\exp \left[ \frac{2(2\rho r)^{1/2}}{\sqrt{N}+\sqrt{N-1}} \right]~.
\end{equation}
The values of $T_n^{(-)}$ for $n<N-1$ are then calculated by the recursion 
formula (\ref{recT}). $T_n^{(-)}$ determined in this way will 
differ from its original definition in (\ref{tn-}) by a multiplicative 
constant. This does not concern us since we are only interested in the 
logarithmic derivative of $\phi_+$, which is independent of the 
constant factor and is readily evaluated by (\ref{logdphi}). 

Leaver's solutions can be applied to compute the logarithmic derivatives 
$f_0(L_{\pm})$ in (\ref{eq:norm2}) and to evaluate the QNMs of the shell 
model of dirty black hole described in \mbox{Sect.\ \ref{sect:shell}}. 

\section{Evaluation of the Perturbation Formulae}\label{app:pert}

For simplicity we focus on the first order perturbation calculation, which is 
given by 

\begin{equation}
  \omega_1=\frac{\langle \phi_0|V_1|\phi_0 \rangle}{2\omega_0 \langle 
\phi_0|\phi_0 \rangle}~. 
\end{equation}
We have discussed the evaluation of the generalized norm 
$\langle \phi_0|\phi_0 \rangle$ in \mbox{Sect.\ \ref{sect:sch}}. It suffices 
to evaluate the generalized matrix element $\langle \phi_0|V_1|\phi_0 
\rangle$. For $V_1$ given by (\ref{eq:shellv1}), the generalized matrix 
element is 

\begin{equation}
  \langle \phi_0|V_1|\phi_0 \rangle=\int_{L_-}^{x_s}V_1(x)\phi_0^2(x)\, dx 
+ \kappa \phi_0(x_s)^2+\left. \phi_0^2(L_-)\frac{\partial}{\partial \mu}
f_-(\omega_0,L_-) \right|_{\mu=0}~,
\label{app:genmat}
\end{equation}
where $L_-$ is any real number smaller than $x_s$. 
It has been argued earlier that this expression
is independent of the choice of $L_-$, and for any finite
$L_-$ the expression is finite.  Thus the entire
perturbation scheme requires no regularization.

However, the last term in (\ref{app:genmat}) 
involves the change of $f_-$ due to the presence of $V_1$.
This calculation can be bypassed by the following
trick.  First, since the entire expression is independent
of $L_-$, we push $L_-$ to $-\infty$ and define 
a function

\begin{equation}
  Z(\omega)=\lim_{L_- \rightarrow -\infty} \left[ \int_{L_-}^{x_s} 
\phi_{0-}^2(\omega,r_*) V_1(r_*)dr_* + \phi_{0-}^2(\omega,L_-)\left. 
\frac{\partial}{\partial \mu}f_-(\omega,L_-)\right|_{\mu=0} \right] 
+ \kappa \phi_{0-}^2(\omega,x_s)~,
\end{equation}
where $\phi_{0-}(\omega,r_*)$ is a function satisfying the unperturbed 
Regge-Wheeler equation with the ingoing wave boundary condition on the 
event horizon.  This expression as a whole is finite for 
every $\omega$, but for 
\mbox{Im$(\omega)>0$}, each term is separately
finite; in fact the surface term vanishes as 
$L_- \rightarrow -\infty$ because the factor 
$\phi_{0-}^2(\omega,L_-) \propto e^{-2i\omega L_-}$ now decays exponentially. 
Evaluating in this domain, we have
 
\begin{equation}
  Z(\omega)= \int_{-\infty}^{x_s} 
\phi_{0-}^2(\omega,r_*) V_1(r_*)\, dr_* \, + \, \kappa 
\phi_{0-}^2(\omega,x_s)\qquad \mbox{for Im$(\omega)>0$~.}
\end{equation}
The prescription is therefore to evaluate in this
domain and analytically continue to $\omega_0$:
\begin{equation}
  \langle \phi_0|V_1|\phi_0 \rangle = Z(\omega_0)=\int_{-\infty}^{x_s}
\phi_0^2(r_*) V_1(r_*) dr_* + \kappa \phi_0^2(x_s)~,
\end{equation}
We note that this merely provides an alternate evaluation
of an expression that was manifestly finite to start with.

The integral can be easily carried out analytically by noticing that the 
potential $V_{SC}(M,x)$ in (\ref{eq:fullpot}) can be expressed as a sum of 
exponentials at negative $x$; the calculation is sketched below
and the details of the straightforward arithmetic can be
found in Ref. \cite{thesis}:
 
\begin{eqnarray}
  V_{SC}(M,x) &=& \bfrac{1}{4M^2}V_{SC}\left(\frac{1}{2},\frac{x}{2M}\right) \\
  V_{SC}(1/2,x) &=& \sum_{k=1}^{\infty} c_k
e^{k\tilde{x}} \qquad \mbox{for }\tilde{x}<0~, 
\end{eqnarray}
where 
\begin{eqnarray}
  c_k &=& l(l+1)\gamma_{3,k}+(1-s^2)\gamma_{4,k} \\
  \gamma_{n,m} &=& e^{-m}\sum_{p=0}^{m} \mu(p,n,m)~.
\end{eqnarray}
Here $\mu(0,n,1)=1-n$, $\mu(1,n,1)=n$, $\mu(p,n,m)=0$ for $p<0$ and $p>m$, 
and 
\begin{equation}
  \mu(p,n,m)\, =\, -\frac{(n+p+m-2)\, \mu(p-1,n,m-1)\, +\, (m-1)\, 
\mu(p,n,m-1)}{m} \qquad \mbox{for $m \geq 2$}~. 
\label{appe:rec1}
\end{equation}
It can be shown that the unperturbed wave function $\phi_0$ can be expressed 
as \cite{expon}
\begin{equation}
  \phi_0(x)=\sum_{k=0}^{\infty} d_k(\omega_0) e^{kx/(2M_0)}~,
\end{equation}
where 
\begin{equation}
d_k(\omega_0)=\frac{1}{k(k-2i\omega_0)}\sum_{m=0}^{k-1}d_k(\omega_0)c_{k-m}~, 
\end{equation}
and $d_0$ is a constant related to the normalization of the wavefunction. 

It is obvious that $V_1$ given by (\ref{eq:shellv1}) can also be 
expressed as a sum of exponentials plus a $\delta$-function. Hence the 
integration in 
(\ref{app:genmat}) can be performed analytically, giving the generalized 
matrix element and the first-order shift $\omega_1$.

\newpage 

\newpage

\newpage

\newpage
\begin{figure}
\epsfig{file=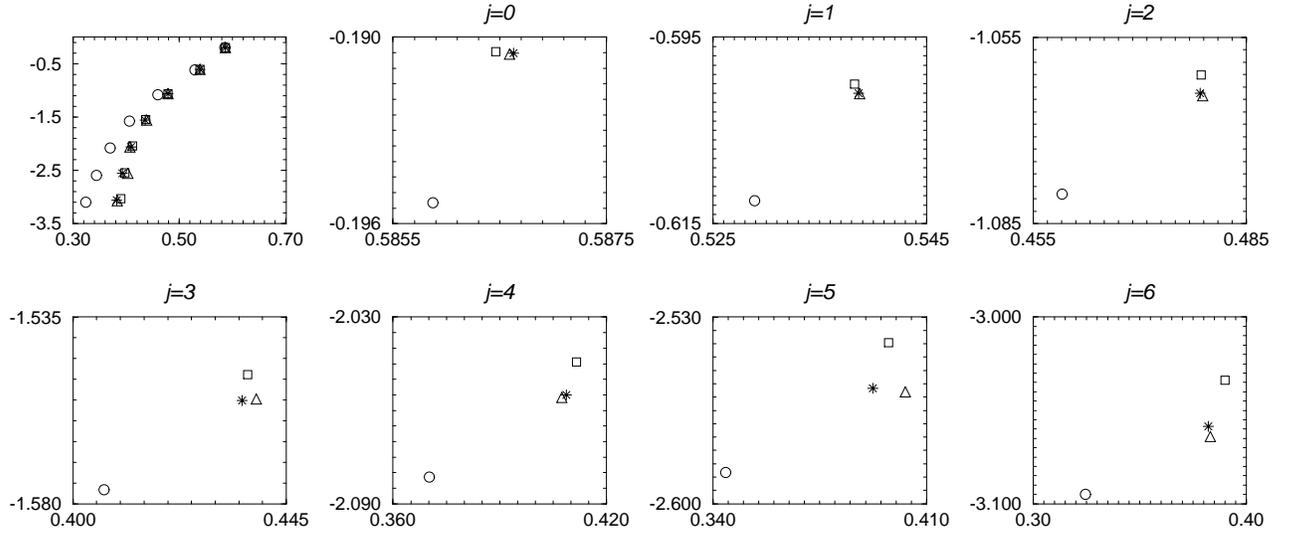,height=17cm,angle=270}
\vskip 2mm
\caption{The perturbed QNM spectrum of a black hole plus shell system for
$l=1$ scalar wave with $\mu=0.02$ and $r_s=2.52 M_a$. The
horizontal axis is \mbox{Re $2M_o\omega$} and the vertical axis is
\mbox{Im $2M_o\omega$}.
The 0th, 1st and 2nd order perturbation result is indicated by circles,
squares and triangles respectively. The exact numerical result is
represented by stars.
The diagram on the upper left shows the distribution of QNMs of the lowest
damping modes (from $j=0$ to $j=6$). The other diagrams are the magnification
of the region around each mode.}
\label{fig:four}
\end{figure}

\newpage

\newpage

\newpage

\newpage

\newpage
\begin{table}
\caption{The normalizing factor $K/(2M)$ (expressed as magnitude and phase) 
for the lowest QNMs of the Schwarzschild black hole with mass $M$.}
\label{tab:one}
\end{table}
\begin{center}
\begin{tabular}{|c||c|r|c|r|c|r|}
 \hline
   & \multicolumn{2}{c|}{$l=0, s=0$} & \multicolumn{2}{c|}{$l=1, s=0$} & 
\multicolumn{2}{c|}{$l=2, s=2$} \\
 \hline
 $j$ & magnitude & phase    & magnitude & phase & magnitude & phase \\
 \hline
 0 & ~~~2.4461 & 36.5$^\circ$ & ~~~4.5155 & 36.7$^\circ$ 
   & ~~~5.5397 & 18.7$^\circ$  \\
 1 & ~193.24~~~~ & 18.4$^\circ$ & ~20.544~ & 68.1$^\circ$
   & ~~~6.9371 & 82.7$^\circ$ \\  
 2 & ~~~6.9935 & 12.9$^\circ$ & ~12.319~ & 22.7$^\circ$ 
   & ~22.770~ & -66.9$^\circ$ \\
 3 & ~894.17~~~~ & -10.2$^\circ$ & ~33.553~ & 10.1$^\circ$
   & ~~~9.4942 & 39.4$^\circ$ \\
 4 & ~12.059~ & 8.4$^\circ$ & ~18.945~ & 20.5$^\circ$
   & ~63.958~ & -89.7$^\circ$ \\
 5 & 2279.0~~~~~ & -10.4$^\circ$ & ~60.316~ & -14.2$^\circ$
   & ~14.358~ & 25.8$^\circ$ \\
 6 & ~17.087~ & 6.7$^\circ$ & ~25.087~ & 26.6$^\circ$
   & ~66.779~ & -100.9$^\circ$ \\
 \hline
\end{tabular}
\end{center}

\newpage
\begin{table}
\caption{The value of $H(x)$ (expressed as magnitude and phase) for
several sample positions at $x=x_1$ ($r=r_1$), for $l=1$ scalar waves
propagating on a Schwarzschild black hole.}
\label{tab:two}
\end{table}
\begin{center}
\begin{tabular}{|c||l|r|l|r|l|r|l|r|l|r|}
 \hline
 & \multicolumn{2}{c|}{$x_1/2M=0$} 
  & \multicolumn{2}{c|}{$x_1/2M=5$} & \multicolumn{2}{c|}{$x_1/2M=10$} 
& \multicolumn{2}{c|}{$r_1/2M=1.02$} & \multicolumn{2}{c|}{$r_1/2M=1.08$} \\
 \hline
 $j$ & magnitude & phase    & magnitude & phase & magnitude & phase
     & magnitude & phase    & magnitude & phase \\
 \hline
 0  & $2.21\times 10^{-1}$ & -36.7$^\circ$  
& $9.18\times 10^{-1}$ & -122.2$^\circ$ & $6.44\times 10^0$ & -147.1$^\circ$
    & $4.16\times 10^{-1}$ & 4.8$^\circ$    & $2.66\times 10^{-1}$ 
& -38.7$^\circ$ \\
 1  & $4.87\times 10^{-2}$ & -68.1$^\circ$  
& $5.23\times 10^1$ & 109.8$^\circ$  & $2.40\times 10^4$ & 52.1$^\circ$ 
   & $2.69\times 10^1$ & -141.1$^\circ$& $2.14\times 10^{-1}$ & 150.9$^\circ$ \\
 2  & $8.12\times 10^{-2}$ & -22.7$^\circ$ 
& $2.88\times 10^3$ & -68.0$^\circ$ & $1.28\times 10^8$ & -165.3$^\circ$
   & $3.14\times 10^1$ & 39.1$^\circ$ & $1.03\times 10^0$ &  -28.5$^\circ$ \\
 3 & $2.98\times 10^{-2}$ & -10.1$^\circ$
& $2.27\times 10^5$ & 96.8$^\circ$  & $1.60\times 10^{12}$ & -30.4$^\circ$
   & $4.20\times 10^2$ & -139.9$^\circ$ & $3.32\times 10^0$ & 156.4$^\circ$ \\
 4 & $5.28\times 10^{-2}$ & -20.5$^\circ$
& $2.28\times 10^7$ & -99.7$^\circ$ & $2.51\times 10^{16}$ & 112.4$^\circ$ 
   & $6.00\times 10^3$ & 40.2$^\circ$ & $1.19\times 10^1$ & -19.2$^\circ$ \\
 5 & $1.66\times 10^{-2}$ & 14.2$^\circ$
& $2.61\times 10^9$ & 65.9$^\circ$ & $4.56\times 10^{20}$ & -97.0$^\circ$ 
 & $9.01\times 10^4$ & -140.2$^\circ$ & $4.27\times 10^1$ & 164.0$^\circ$ \\
 6 & $3.99\times 10^{-2}$ & -18.0$^\circ$
& $3.22\times 10^{11}$ & -126.3$^\circ$ & $8.91\times 10^{24}$ & 59.6$^\circ$
  & $1.40\times 10^6$  & 39.4$^\circ$ & $1.57\times 10^2$ & 13.6$^\circ$ \\
 \hline
\end{tabular}
\end{center}


\begin{references}
\bibitem{fn1}
The ``extraneous'' factor of $2\omega_0$ (compared to the
usual result for the Schr\"{o}dinger equation) occurs because
the eigenvalue is $\omega^2$, 
and $\Delta(\omega^2) = 2\omega \Delta\omega$,
\bibitem{norm} 
For the particular case of potentials with both $V_0$ and
$V_1$ having finite support, the generalized norm has been introduced
previously both for the wave equation \cite{waveeq}, the
Schr\"{o}dinger equation \cite{schrod} and the Klein-Gordon equation
\cite{ching-comp}, and its property discussed.  An equivalent form
first given by Zeldovich \cite{zel} did not have the surface terms,
but instead required a process of regularization; it is less
convenient when actual evaluation (especially numerical evaluation) is
called for.  The present result, in the more general form
(\ref{eq:norm}), is however applicable to potentials that extend to
infinity, including, for example, the important Regge-Wheeler
potential (\ref{eq:schw}) to be discussed below.
\end{references}
\end{document}